\documentclass[a4wide]{article}

\usepackage{a4wide}
\usepackage{comment}
\usepackage{pifont}% http://ctan.org/pkg/pifont
\usepackage{array}
\usepackage{booktabs}
\newcommand{\git}{Git }
\newcommand{\github}{GitHub }
\newcommand{\mycheckmark}{\ding{51}}%
\newcommand{\cross}{\ding{53}}%
\usepackage{balance}
\usepackage{hyperref}
\usepackage{amsmath}
\usepackage{cleveref}
\usepackage{enumitem}
\usepackage{xspace}

\usepackage{url}
\usepackage{tablefootnote}
\usepackage{listings}
\usepackage{xcolor}
\usepackage{todonotes}
\usepackage[numbers]{natbib}

\colorlet{punct}{red!60!black}
\definecolor{background}{HTML}{EEEEEE}
\definecolor{delim}{RGB}{20,105,176}
\colorlet{numb}{magenta!60!black}

\lstdefinelanguage{json}{
    basicstyle=\normalfont\ttfamily,
    numbers=left,
    numberstyle=\scriptsize,
    stepnumber=1,
    numbersep=8pt,
    showstringspaces=false,
    breaklines=true,
    frame=lines,
    backgroundcolor=\color{background},
    literate=
     *{:}{{{\color{punct}{:}}}}{1}
      {,}{{{\color{punct}{,}}}}{1}
      {\{}{{{\color{delim}{\{}}}}{1}
      {\}}{{{\color{delim}{\}}}}}{1}
      {[}{{{\color{delim}{[}}}}{1}
      {]}{{{\color{delim}{]}}}}{1},
}

\newcommand{\eg}{e.g.\xspace}

\usepackage{tikz}

\def\halfcheckmark{\tikz\draw[scale=0.4,fill=black](0,.35) -- (.25,0) -- (1,.7) -- (.25,.15) -- cycle (0.75,0.2) -- (0.77,0.2)  -- (0.4,0.7) -- cycle;}

\newcommand{\myciteauthoryear}[1]{%
  \citeauthor{#1} (\citeyear{#1})%
}

\title{A Backend Platform for Supporting the Reproducibility of Computational Experiments} % Article title, use manual lines breaks (\\) to beautify the layout

\author{%
	Lázaro Costa\textsuperscript{1,2}, Susana Barbosa\textsuperscript{2}, Jácome Cunha\textsuperscript{1,2}\\
	lazaro@fe.up.pt, susana.a.barbosa@inesctec.pt, jacome@fe.up.pt}

\date{
\footnotesize\textsuperscript{\textbf{1}} Faculty of Engineering, University of Porto, Portugal\\ \textsuperscript{\textbf{2} INESC TEC, Portugal}}

\begin{document}

\maketitle % Output the title section

% Full-width abstract
%\renewcommand{\maketitlehookd}
	\begin{abstract}
	%	\noindent
In recent years, the research community has raised serious questions about the reproducibility of scientific work. In particular, since many studies include some kind of computing work (for instance, executing an algorithm encoded using some programming language), reproducibility is also a technological challenge, not only in computer science, but in most research domains.

Replicability and computational reproducibility are not easy to achieve, not only because researchers have diverse proficiency in computing technologies, but also because of the variety of computational environments that can be used. Indeed, it is challenging to recreate the same environment using the same frameworks, code, %data sources,
programming languages, dependencies, and so on.

In this work, we propose an Integrated Development Environment allowing the share, configuration, packaging and execution of an experiment by setting the code and data used and defining the programming languages, code, dependencies, databases, or commands to execute to achieve consistent results for each experiment. After the initial creation and configuration, the experiment can be executed any number of times, always producing exactly the same results. Furthermore, it allows the execution of the experiment by using a different associated dataset, and it can be possible to verify the reproducibility and replicability of the results. %This allows the creation of a reproducible pack that can be re-executed by anyone on any other computer.
Our platform aims to allow researchers in any field to create a reproducibility package for their science that can be re-executed on any other computer.

To evaluate our platform, we used it to reproduce 25 experiments extracted from published papers. We have been able to successfully reproduce 20 (80\%) of these experiments achieving the results reported in such works with minimum effort, thus showing that our approach is effective.

% Please include a maximum of seven keywords
	\end{abstract}

%----------------------------------------------------------------------------------------
%	ARTICLE CONTENTS
%----------------------------------------------------------------------------------------

\section{Introduction}\label{sec:intro}
Reproducibility is the ability to replicate the results of a study using the original methods, materials, and data. It is paramount for science and is widely considered to be a precondition of scientific advancement \cite{Bush2020, costa2022platform}.
%On the other hand, 
Replicability is the capacity to obtain the same results using new methods, materials, and data and/or consistent conditions with the original study~\cite{Bush2020, costa2022platform}. %In our work, we focus on reproducibility.

In recent years, many concerns have been raised related to the reproducibility of today's computational work. A significant amount of scientific research is nowadays supported by some kind of computational experiment. Thus, reproducible research is now more than ever a technological challenge in many research domains.

Computational reproducibility should be easily achievable since it is possible to specify detailed information about the computational environment being executed. This should enable the reproduction of the computation environment in the same way. But in practice, this does not happen (see \cref{sec:survey}). The complexity of today's software makes the reproducibility process more difficult~\cite{Ivie2018}.
%The complexity of today’s software and hardware makes this procedure very hard for humans and machines to understand it accurately~\cite{Ivie2018}.

Research in many fields (\eg, chemistry, climate change, biology) is supported, in several studies, by computational experiments~\cite{Ivie2018}.
However, many of these researchers are not experts in software development. 
Thus, although they may be able to create the necessary code for their experiments (\eg, scripts in Python or R), they may find it difficult to create reproducible computing experiments. In fact, this is even the case for computer science conferences (\cref{sec:survey}).

To keep research reproducible, it is essential to investigate, perform, analyze, and rerun the experiments associated with that research in order to obtain corresponding results as before. For this to be possible, when working with a computational experiment, it is necessary to rebuild the original computing environment using the original code and its dependencies, the same data, and the same script-based workflow~\cite{Nust2021,Brunsdon2021}.

Describing all the dependencies used and which versions are used is not a trivial task. 
Package management tools of each programming language (PL), such as pip\footnote{\url{https://pypi.org/project/pip/}} for Python or npm\footnote{\url{https://www.npmjs.com}} for JavaScript, have integrated commands which can build a file with the dependencies needed for an experiment.
However, these files do not have all the information to enable computational reproducibility. Moreover, researchers without a computer science background may not know how to use such tools~\cite{Ivie2018}.
There are tools to aid researchers in making their computational experiments reproducible such as 
Open Lab~\cite{Shevchenko2022}, SciInc~\cite{Youngdahl2019}, CDE-SP~\cite{Pham2015}, Encapsulator~\cite{Pasquier2018}, 
A Framework~\cite{Haiyan2015}, CARE~\cite{Janin2014}, 
VisTrails~\cite{Freire2012}, Binder~\cite{Matthias2018binder}, 
RenkuLab~\cite{krieger2021repeatable}, 
Whole Tale~\cite{Brinckman2019}, ReproZip~\cite{Chirigati2013,Chirigati2016,Chirigati2016_colaborative}, or
RunMyCode~\cite{Stodden2012}.
However, these approaches are limited to a set of PLs, do not allow the use of more than one PL per project, do not allow the use of databases, or do not allow researchers to specify which commands should be executed to achieve the experiment results.
The Association for Computing Machinery (ACM) has motivated researchers to build a reproducibility package by creating a virtual machine. However, again this is not easy for many researchers.
%the set up of the experimental environment, including the validations of the reproducibility.% and are not possible to search and reuse datasets.
%In many cases the platforms are not very user-friendly, \eg, requiring users to know how to use \git repositories. 
In \cref{sec:related} we further discuss related work.

To assess the use of reproducibility tools, we defined the following research question (RQ), which we answer in \cref{sec:survey}:\\

\noindent
\textbf{RQ1: What is the adoption of reproducibility tools among researchers?}

To answer this RQ we designed and distributed an online survey.
Although the aforementioned approaches allow the reproducibility of experiments, just one of the 38 researchers answering our survey acknowledged to have used one of these platforms to save and reproduce computational experiments. 

Thus, with our work, we mainly seek to answer the following RQ (\cref{sec:Methodology}):\\

\noindent
\textbf{RQ2: To which extent can we propose a platform that helps to make computational experiments from different fields reproducible?}

To answer this RQ, we propose an Integrated Development Environment (IDE) that allows the share, creation, configuration, (re-) execution and packaging of computational experiments, % and process of computational experiments
which also includes the validation of reproducibility and replicability of the results. %Researchers can compare the results achieved with the results of previous runs by using the associated data or different data.
Researchers need to associate a dataset to each experiment, however, it is possible to execute using a different dataset. In this way, researchers can compare the results achieved with the results of previous runs by using the associated data or different data.

For now, we focus our work on the backend of the platform, not providing a user interface (which we plan to design and build in future work).
Our platform is developed to support a variety of experiments, including Artificial Intelligent (AI) experiments. For these experiments, we make researchers aware of the importance of making the experience reproducible, helping them associate the possible seed used with the project. In addition, 
Furthermore, our approach can be used to create a capsule of the research experiment containing the code and its dependencies, the data and databases, the process, and the commands to achieve the experiment results. 
Our main goal is to support the researchers in developing, executing and sharing the code they produce for their experiments, allowing them to manage the data and code associated with that experiment, without the need to manage dependencies, compilers, and other variables.
The proposed platform can handle many PLs, databases, and other data files. It can also automatically infer information from the experiment, such as PLs and dependencies, creating the proper environment for the execution. The researcher can then easily export a reproducibility package that can be executed on any computer just by double-clicking a single file, making the experiment easily sharable and re-executable.
%With the development of our approach, we intend to improve the data reusability and the reproducibility of experiments.

To evaluate our platform, we selected a set of experiments included in already published papers and tried to recreate them using our platform. In particular, we seek to answer the following RQ (\cref{sec:experimental}):\\

\noindent
\textbf{RQ3: What kinds of experiments can supported?}

To answer this RQ, we collected experiments from related work, from major conferences, and some others from Zenodo so we could have a variety of experiments. We collected 31 experiments from fields such as computer science, medicine, or even climate change science. We were, in general, successful in reproducing the experiments, achieving the same published results, and were able to create a reproducibility package that anyone can re-execute just by executing a single file.

Based on these RQs and our answers to them, the main contributions of this work are:

\begin{itemize}
    \item a set of principles that a reproducibility platform should follow (\cref{sec:challengesInReproducibility});
    
    \item an initial overview of the perspectives of researchers about reproducibility difficulties and tool usage (\cref{sec:survey});

    \item the backend of a reproducibility platform tackling the reproducibility principles and the researchers' perspectives that can cope with experiments from very different scientific fields (\cref{sec:Methodology});

    \item a list of several reproducibility packages for existing scientific publications (\cref{sec:experimental});
\end{itemize}

Before we introduce our platform, in \cref{sec:challengesInReproducibility} we discuss in more detail the challenges of reproducing computational experiments, and we define a set of principles a platform supporting computational reproducibility must follow to be successful. This may help future researchers proposing other approaches for reproducibility.

%For decades, the topic has intrigued researchers working with large-scale survey data, archivists at institutional repositories, and individuals who were frustrated with unsuccessful attempts to obtain other researchers' data.
%Fear (2013) asserts that this tradition of sharing and reusing data in the social sciences is due to the nature of social research, which often requires large amounts of unique data collected over time~\cite{fear2013measuring, YOON2017224}

%In principle, it should be possible to specify a computation to sufficient detail that anyone should be able to reproduce it exactly. But in practice, there are fundam        ental, technical, and social barriers to doing so. The many objectives and meanings of reproducibility are discussed within the context of scientific computing. Technical barriers to reproducibility are described, extant approaches surveyed, and open areas of research are identified~\cite{Ivie2018}.

\section{Challenges in Reproducibility}\label{sec:challengesInReproducibility}
The digital transformation of science has made available new publication channels for research data. These publications come from specialized data repositories, large centralized or decentralized platform infrastructures, or commercial application providers to traditional self-hosted web server downloads.
Some examples of these platforms are 
CKAN\footnote{\url{https://ckan.org/}}, DSpace\footnote{\url{https://dspace.lyrasis.org/}}, EUDAT\footnote{\url{https://www.eudat.eu/}}, 
\github\footnote{\url{https://github.com/}}, 
the Harvard Dataverse\footnote{\url{https://dataverse.harvard.edu/}}, Zenodo\footnote{\url{https://zenodo.org/}}, and derived systems.

Nowadays, these platforms provide ways to upload and publish research datasets with an institutional, disciplinary focus, or general purpose. Such platforms also need to fill the needs of different stakeholders in a variety of usage scenarios~\cite{Langer2019, kim2018internet}.

Although these platform help spread science, they are not enough to make it reproducible.
Researchers need to expose their results, but also other artefacts, such as code and data, and explain how to use them to achieve the published results. However, these platforms do not provide means to make the code easily reproducible.

Based on previous work, we summarize a set of challenges a researcher needs to take into consideration when intending to make an experiment easily reproducible:

\begin{description}[leftmargin=8pt]
  \item[\textbf{C1: Software}]
    In the research process, the software has a high impact on the reproducibility of experiments \cite{Sarah2017}.

    When someone tries to reproduce an experiment but uses the wrong software, the wrong version, or even the wrong dependencies, the experiment may produce different results from the original one. It is known as Code Rot~\cite{boettiger2015}. The management of the software to use is difficult as even newer versions of software that claim to be compatible with the previous ones, may have unintended differences that can affect reproducibility. To allow reproducibility, all the information related to the software should be identified to ensure a consistent behavior~\cite{Ivie2018}.
  
    The term Dependency Hell refers to a common issue in software development where multiple packages share a dependency, but depend on different and incompatible versions of that dependency. As only one version of a dependency is allowed in a project's environment, resolving these conflicts and finding a compatible solution can be challenging~\cite{Pham2015}.

    Another potential problem with reproducibility is when the libraries are unavailable. To get these libraries, it is necessary to find them which hinders reproducibility.  

    Software is the tool that helps the researcher in the reproducible process (create, configure, run, encapsulate and re-run), however, it is not easy to find easy-to-use and complete solutions~\cite{Freire2012_b, Zuduo2021}.

   \item[\textbf{C2: Command}]
    It is crucial to make available the commands to execute the code, along with the file to execute and all the values of the parameters needed to reproduce the experiment.
    Some researchers prefer not to use parameters on the command, but this approach makes it difficult for those who intend to extend or explore the research. So using parametrization is an important method for extension, as it could give both the original researcher and other researchers the ability to easily extend the approach, explore the parameter space, and validate results~\cite {Ivie2018}.
  
    However, using a single command with too many parameters can be very confusing for other researchers to extend the experiment. To make the extension exploration and reproducibility easier, it is better to simplify the command and break it into smaller commands and smaller workflows. In the end, it is necessary to explain what each command does and organise them in a general workflow~\cite{aslst2002workflow,liu2015survey, Ivie2018}.
  
    Furthermore, the expected result should be available for comparison, because there is no guarantee that the same command applied by a different user on a different machine will get the same result. This can happen due to the differences in the dependencies installed, the software, the data used, and the dependencies implicitly provided by the system~\cite{Ivie2018}.
  
  \item[\textbf{C3: Data}]
    Making available all the data is another important element to allow reproducibility~\cite{David2019}.
    Most researchers use some kind of input as starting data as well as the final generated results of their experiments. These inputs sometimes are in remote locations, so security or privacy issues can occur during the data download. 
  
    Some data files may contain private information, that is normally omitted from the publication, but without it, the experiment is not reproducible. Receiving this information as an input parameter can identify the need for the information without publishing the delicate data, enhancing reproducibility~\cite{Ivie2018}.

    Furthermore, researchers intend to find and reuse existent data, but the existing platforms and approaches do not include sufficient metadata (\eg, title, authors, version, PLs, software description, Uniform Resource Identifier/DOI and execution requirements) to find and reuse data~\cite{Victoria2016}.
  
   \item[\textbf{C4: Hardware}] 
    Enabling reproducibility using the same performance, scalability, and efficiency is another challenge because the same hardware should be used (\eg, the same computing power such as RAM, CPU, GPU, and storage).
     However, in this work, we will not take into consideration experiments that may have hardware requirements.
  \end{description}

%In the research domain, some principles are crucial during the research process to enable the creation of computational environments, execution of experiments, validations of the results and sharing of the process previously built. 

From the challenges C1, C2, and C3 we just discussed, we identify a set of principles a reproducibility platform should follow. We also introduce principles that are orthogonal to the challenges.

\begin{description}[leftmargin=1em]
\item[From C1]  \ \\  \vspace{-.4cm}
    \begin{description}[leftmargin=0em]
       \item[\textbf{P1}] Such a reproducibility platform should allow researchers to define the PLs used in the experiment, the compilers, the dependencies, and the libraries, as well as the correct version of each.

       \item[\textbf{P2}] It should allow saving all the dependencies needed to execute the experiment together with the code.
    \end{description}

\item[From C2] \ \\ \vspace{-.4cm}
    \begin{description}[leftmargin=0em]
        \item[\textbf{P3}] The platform should support the configuration of the commands for the execution of the experiment.

        \item[\textbf{P4}] It should allow the validation of the computational reproducibility using the results of previously performed experiments as a means of comparison.
    \end{description}

\item[From C3] \ \\ \vspace{-.4cm}
    \begin{description}[leftmargin=0em]
        \item[\textbf{P5}] Such a platform should provide mechanisms that allow researchers to manage the data of the experiment.
    \end{description}

\item[Orthogonal principals] \ \\ \vspace{-.4cm}
    \begin{description}[leftmargin=0em]
        \item[\textbf{P6}] It should be able to deal with the diversity of PLs (and their dependencies) used by researchers.

        \item[\textbf{P7}] It should support all types of experiments, from the execution of simple scripts to the execution of more complex experiments (\eg, involving databases or several PLs).
    
        \item[\textbf{P8}] The platform should support researchers with
        little knowledge of computing. 
        
        \item[\textbf{P9}] It should enable the sharing of the experience easily.

        \item[\textbf{P10}] It should 
        facilitate the re-execution of the experiment even for users outside the research domain.
    \end{description}

\end{description}

In this section, we have synthesized the challenges already identified in the literature. However, we also want to have some more insights from the researchers themselves. Thus, in the following section, we present a small survey about the views of researchers on reproducibility.

\section{Researchers' Experience with Reproducibility}\label{sec:survey}
We performed a survey\footnote{Available at \url{https://docs.google.com/forms/d/e/1FAIpQLScUtalm-7X0BwjKbNtpam5WDywQ8wWVTzfO3sfE8K0yA0c1EQ/viewform}.} in which the main goal was to gather some information related to the familiarity of researchers with the concept, the challenges, requirements, and the main difficulties of computational reproducibility. Furthermore, it is important to know which platforms researchers use/know to reproduce experiments. Note that this survey is not intended to be complete, but rather to give us a more informed view about researchers' perspectives on reproducibility.

We made available this survey to all the authors of the 2022 edition of the International Conference on Software Engineering since we used some of these works in our evaluation (see \cref{sec:experimental}). From the 700 contacted authors, we got 38 responses. Moreover, we spread the same survey on social media platforms (Twitter, Facebook, and LinkedIn) through public posts, and also distributed the survey through our contacts via email, for which we got 38 responses too.

Our sample is composed mostly of Ph.D. students (46\%), followed by faculty members (33\%) and researchers (11\%). They are from Portugal (46\%), China (16\%), and the USA (13\%). Furthermore, 90\% of the respondents have more than two years of research experience.

The ``computational reproducibility'' term is familiar to 84\% of the surveyed, and 79\% tried to reproduce experiments performed by others. However, most of them, 66\%, had difficulties in doing so. %In \cref{fig:difficulties} we can see a chart that contains the perceived level of difficulty of researchers when trying to reproduce experiments (the higher the level the more difficulties the researchers felt).

%\begin{figure}[!htb]
%\centerline{\includegraphics[width=\columnwidth]{images/difficulties.png}}
%\caption{Researchers' level of difficulties when trying to reproduce experiments}
%\label{fig:difficulties}
%\end{figure}

As we can see, only a few researchers did not find it difficult to reproduce others' works.

The main challenges in computational reproducibility found in this survey are ``Missing information to execute the experiment'' (34\%), ``Data not available'' (25\%), ``Code not available'' (24\%), and ``Installation not successful'' (17\%).

Only 11\% of the surveyed know a reproducibility platform, namely Binder, while all the others do not know any.
Only 8\% of the surveyed reproduce their experiments using docker\footnote{Docker is a platform designed to help developers build, share, and run modern applications. Docker uses service virtualization at the operating system level (https://www.docker.com/).}, 8\% of the surveyed publish the code on Zenodo, and 5\% on \github.

Finally, we asked what requirements were considered necessary to allow computational reproducibility. The answers were ``Make available the data'' (29\%), ``Make available the code'' (28\%), ``Make available the description of all the dependencies used and its versions'' (24\%), and ``Make available the PL used and its version'' (18\%).

To conclude, the answers to this survey show us that the term computational reproducibility is a term well-known in the research community. However, researchers usually have difficulties when trying to reproduce experiments performed by others. %We get the main challenges and the main requirements that they considered to hinder computational reproducibility.
Although with this small survey, we do not have a complete view of the researchers' perspectives on reproducibility, we can conclude that, although there are already some proposals to aid in reproducing computational experiments (see \cref{sec:related}), the researchers that answered this survey still do not use them at scale, which also motivates our own work.

%Based on the principles established in \cref{sec:challengesInReproducibility} and on the results from the survey, in the next section, we detail our efforts to design and implement a reproducibility platform that can cope with the principles and the researchers' concerns.

\section{Design and Implementation of a Reproducibility Platform}\label{sec:Methodology}
%In the research domain, some approaches allow reproducibility. However, they are not designed to support the research process of all researchers, due to the different levels of knowledge in computer science.

Based on the principles defined in \cref{sec:challengesInReproducibility} and the data collected from the researchers described in \cref{sec:survey} in this section, we present our design and implementation of a backend of a platform that gives support to the creation and execution of computational experiments.
%Notice we focus on reproducibility and, in particular, experiments executed in a computational environment. 
Notice we focus our work on the software environment and assume the hardware does not influence the outcome of the experiments (although in some kinds of experiments that are actually intended, we will not consider such cases).
In this work, we do not provide a user interface, which we plan to design and build in future work. 
%We plan that our approach is to be able to deal with the diversity of PLs, manage the code and the data of the experiment and capable to assist all types of experiments. Furthermore, our goal is to support the researcher in the configuration of the execution of the project and enable the validation of the computation reproducibility. With this, we plan that the experiments can be executed any number of times, producing always exactly the same results. 

\begin{figure}[!htb]
\centerline{\includegraphics[width=\columnwidth]{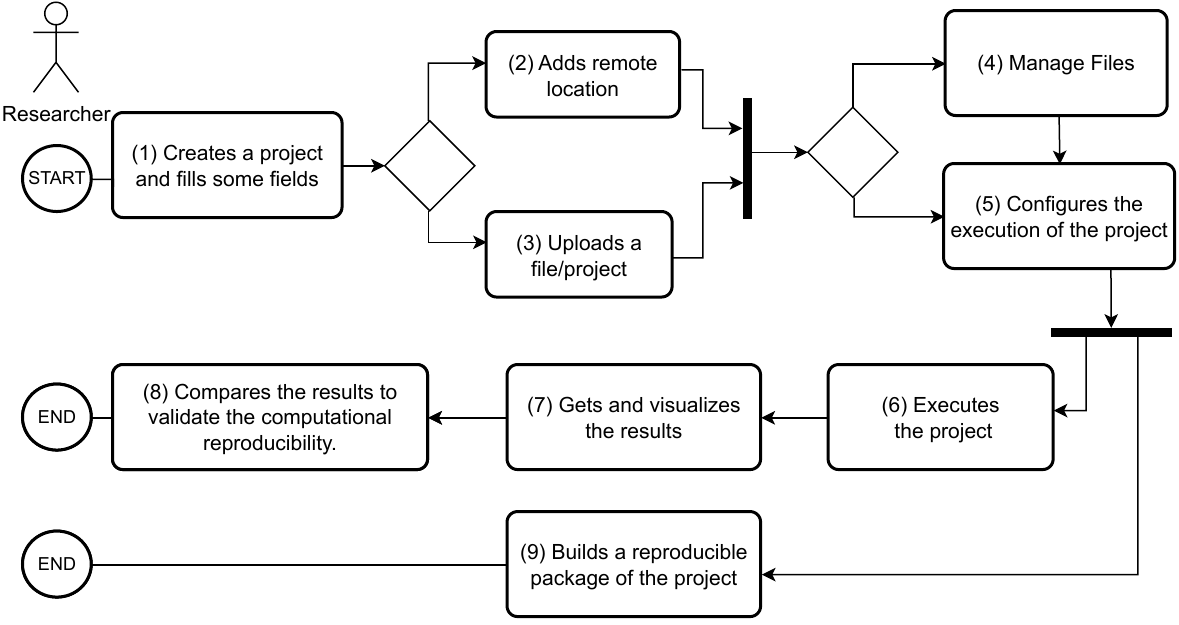}}
\caption{The activity diagram of a researcher using our approach}
\label{fig:activityDiagram}
\end{figure}

%With our work, we do not intend to change researchers' everyday work but help when they are ready for sharing their findings. Thus, our methodology does not intend to aid researchers in designing, preparing, or executing their experiments. Rather, we intend to provide support after the work is concluded and ready for publication. As researchers usually publish their works in written documents, we aim to help them (in parallel) prepare a reproducibility package that accompanies the paper. This can help reviewers to easily re-execute the experiments, but also any other person that intends to do so.
The methodology that researchers should follow to create a reproducibility package using our platform can be formulated in the following chronological order:
\begin{enumerate}[leftmargin=16pt]
    \item Researchers create a new project, choose the type of the project and fill in some fields (\eg, project name, description, etc.). For now, we only support the execution of scripts with a database integrated and AI experiments. However, the latter is necessary to make researchers aware of the reproducibility of the experiment, so we force researchers to associate the seeds used in the code with the project (Step 1 in \cref{fig:activityDiagram});
    
    \item They upload the experiment files (\eg, code and datasets) or reuse a pre-existent project by providing its remote location. This functionality enables the uploading of individual or zip files, which can contain all the needed files of the experiment (Steps 2 and 3 in \cref{fig:activityDiagram}). 
   
    \item They can manage the files within the platform (\eg, create some missing files/folders, edit and delete some unnecessary files/folders loaded by mistake or no longer needed) (Step 4 in \cref{fig:activityDiagram});
    
    \item They configure the execution of the project and associate the dataset of that execution (\eg, choose a specific version of the PL and configure the databases) (Step 5 in \cref{fig:activityDiagram});
    This functionality can infer important information such as the PLs, dependencies and their version, however, the researcher must validate that information.    
    In addition, it allows the configuration of AI experiments. These experiments cannot be reproducible if researchers do not define the seeds used. To try to solve this problem, we are making researchers aware that they need to define the seeds used in the project. 
    Furthermore, the researcher can provide more information (\eg, commands to build the project, and the configurations of possible databases) to generate a container specification and build a container image of the experiment. In addition, researchers need to associate a dataset to each experiment, allowing them to execute the experiment using a different associated dataset. In this way, it is possible to verify the computational reproducibility and replicability.
    After the platform concludes this step, our approach will execute the experiment 2 times to verify the computational reproducibility (Step 5 in \cref{fig:activityDiagram});
    
    \item Researchers execute the experiment using the dataset associated or a different dataset. This functionality enables the reusing of container images, generated in the previous step, to start a new execution. In addition, it allows adding a pre-existing container image to add a database to the run. All the executions which use the same container image reuse the same computational environment to execute the experiment, however, if needed, the user can use the same computational environment but using a different dataset. Furthermore, it is possible to specify which commands one would like to run (Step 6 in \cref{fig:activityDiagram}); 
    
    \item Researchers get and visualize the results (Step 7 in \cref{fig:activityDiagram});
    
    \item Validate the computational reproducibility or the computational replicability.%They compare the results with the results of the other executions
    Researchers can define the location of the output (\eg, console output, a file, etc.), however, if it is not defined, we will compare all files changed and the console output 
    (Step 8 in \cref{fig:activityDiagram});
    
    \item Researchers build a reproducible package of the experiment. It allows building a reproducible package, which contains all the information needed to execute the experiment, that can be run on Windows, Mac OS or Unix operating systems. It reuses the container image previously generated and adds all the necessary information, including the commands executed when running the experiment (Step 9 in \cref{fig:activityDiagram}).
\end{enumerate}

 \cref{fig:activityDiagram} presents an activity diagram formalizing this process.

\subsection{Architecture}\label{sec:implementationArchitecture}

\begin{figure}[!htb]
\centerline{\includegraphics[width=\columnwidth]{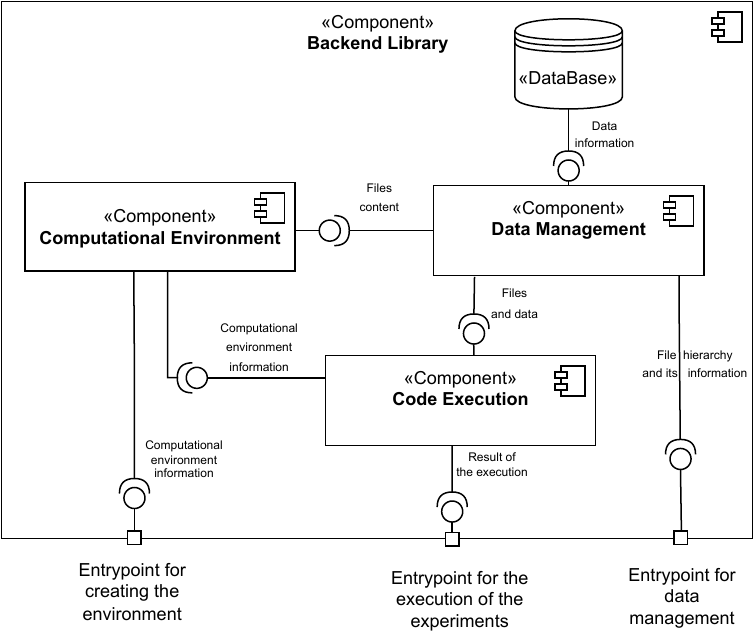}}
\caption{The components' diagram of our platform}
\label{fig:components-diagram}
\end{figure}

Our platform has three main components, namely ``Data Management'', ``Computational Environment'', and ``Code Execution''. Furthermore, we have a database, which saves some required information.
\cref{fig:components-diagram} represents the components' diagram of this approach. In the following, we explain in detail each one.

\subsubsection{Data Management}\label{sec:data}
This component is responsible for managing all the data of the platform, such as adding, editing and deleting files and folders. The researcher can use this component to upload all the necessary files on a project (\eg, code, datasets, etc.), and the Data Management Component saves on the database all the information of each file/folder.

%We implemented the functionalities of
This component complies with the principle P5 from \cref{sec:challengesInReproducibility}.

\subsubsection{Computational Environment}\label{sec:environment}
The function of this component is to build a computing environment able to deal with the diversity of development stacks (\eg, a variety of PLs, dependencies and, if needed, databases integrated) used by researchers and make that information available to the Code Execution Component (\cref{sec:code}) which executes the experiment.

The diversity of available computing environments is a challenge in general software development and is being tackled by using \textit{containers}~\cite{clyburne2019computational, Amit2020}, a technology that allows automatically building an environment based on a specification.
A specification file for a container, also known as a container specification, is a document, like a template, that outlines the configuration and requirements of a containerized application or service. It describes the resources, environment, dependencies, and other necessary information for the container to run consistently across different environments~\cite{clyburne2019computational, Amit2020}. One of the most used containerization technologies is Docker\footnote{\url{www.docker.com}}~\cite{Amit2020}.
%
%The specification file used to execute code in a Docker container is called dockerfile or docker image. These specifications work as a set of instructions, like a template, to build a Docker container~\cite{rad2017introduction, Amit2020}. 
%
For instance, one may define a specification of an environment with Python 3, PostgreSQL 15.1 and Apache 2.4. By using the same specification, each member of a team works in the exact same environment as all the others. We build on this same principle to allow researchers to reproduce their experiments in a systematic way as the Computational Environment component is based on containerization.

This component complies with the P1, P2, P6 and P7 principles from \cref{sec:challengesInReproducibility}.

%\begin{description}[leftmargin=8pt]
 %   \item [\textbf{Build a container image}] %The user can choose to configure the computational environment of the experiment, namely, setting the PLs and their version, dependencies to add, and commands to execute. %The response to this is the unique identifier of the docker image created.
    %The PLs, which the user intends to have in the building of the computational environment, is the only mandatory parameter. If the non-mandatory parameters are missing, the latest version of the PL will be used and no more settings will be added. 
    
    %The creation of the database is optional, but it can be done using this request too. For that to be possible, the database to use is a required parameter.
    %To configure the database, parameters such as username, user password, the name of the database, the root username and the root password, should be filled in these parameters (Step 4 in \cref{fig:activityDiagram}).% The response is the unique identifier of the docker image created for the database.
%\end{description}

\subsubsection{Code Execution}\label{sec:code}
This component is responsible for the execution of the experiment code/scripts, validating the reproducibility or replicability of results and building a reproducible package with all the information needed to re-execute the experience in the same way.

This component complies with the P3, P4, P9 and P10 principles from \cref{sec:challengesInReproducibility}.

%\begin{description}[leftmargin=8pt]
 %   \item[\textbf{Run an experiment}] 
  %  \item[\textbf{Validate the reproducibility or replicability of results}]
    %\item[\textbf{Build a reproducible package}] 
    %where the only requirement is to have Docker installed and running. This package can then simply be clicked and the experiment is executed (Step 8 in \cref{fig:activityDiagram}). 
%\end{description}

\subsection{Implementation}
In this section, we describe some relevant details of our implementation.
We implement our approach in Python\footnote{\url{https://www.python.org/}} and Node.js\footnote{\url{https://nodejs.org/en/}} and use a Neo4j\footnote{\url{https://neo4j.com/}} NoSql database, as well as Docker Hub.

The Data Management component is implemented in Node.js. We are using the Express\footnote{\url{https://expressjs.com/}} framework for communication. The information of all the files and folders of an experiment are saved on a Neo4j database.

%\subsubsection{Node Server}\label{subsec:NodeServer}
%The Data Management component is implemented in Node.js. We are using the Express\footnote{\url{https://expressjs.com/}} framework for communication. The information of all the files and folders of an experiment are saved on a Neo4j database.

%\subsubsection{Python Server}\label{subsec:PythonServer}
The package responsible for the creation, configuration, execution and packaging of experiments is implemented in Python. We are using the Flask\footnote{\url{https://flask.palletsprojects.com/en/2.2.x/}} framework for communication and we are also using Docker Hub as the containerization technology.

%to build a docker image of the experiment and enable the execution of containers. 

%The Python micro-service (\cref{fig:components-diagram}) has two integrated components: ``Computational Environment'' and ``Code execution'' components.

Our approach can build a variety of computing environments needed for researchers to execute their experiments. Our approach already supports a variety of PLs, namely C++, Perl, R, Java (maven), Python, Unix Shell and Jupyter Notebook. 
We have implemented a heuristics-based algorithm that allows the platform to infer the PLs of the project through file extensions. Moreover, our approach can also infer the dependencies to be installed, by constructing the project dependency graph.

The platform infers all the information needed to build the container specification, as the technology is Docker, this file is called dockerfile.

The platform performs all the steps of \cref{fig:activityDiagram}, including building a reproducible package of the experiment. However, the execution of this reproducible package, which contains all the information needed to execute the experiment, is on the researcher's Windows, Mac Os or Unix operating system. The only requirement is to have Docker installed and running. This package can then simply be clicked and the experiment is executed.

It is easy to add another PL. For that to be possible, it is only necessary to do some updates to get a docker image of the wanted language from Docker Hub\footnote{Docker Hub is the world's largest container image repository containing open source projects that are made available in containers (\url{https://hub.docker.com/}).} and build a dockerfile with that information. However, we do not expect this to be done by researchers, but instead by a maintenance team of the tool.

Our approach supports databases, and it allows researchers to configure their setup. The databases currently supported are Mongo, SQLite MySql and PostgreSQL.
As with PLs, it is easy to support another database technology.

\section{Evaluation}\label{sec:experimental}
%\subsection{Evaluation experiment}
We build an evaluation experiment to measure the capabilities of the platform. The main goal of this experiment is to reproduce a variety of experiments, from a simple script to experiments with integrated databases. We also considered experiments from different scientific fields. For that to be possible, we need to recreate the same environment using the same frameworks, code, data sources, the same PLs, dependencies, and the same version of the dependencies. After that, we execute the experiment and reproduce the results claims on the publication. 

%It is to be considered that only the API of the application is being used in this evaluation experiment.

\subsection{Setup}
We started by collecting experiments already published in scientific conferences. To do so, we used two different computer science research communities: software engineering through the conference \textit{IEEE/ACM International Conference on Software Engineering} (ICSE), and databases through the conference \textit{International Conference on Very Large Databases} (VLDB), three very well-known scientific events in their respective area. 
%We chose computer science conferences in the expectation that such works would pose the most complex challenges when compared to other research areas where one would expect the experiments would be simpler. 
We analysed the first 30 articles of VLDB 2021\footnote{\url{http://vldb.org/pvldb/volumes/15/}} and the first 50 articles of ICSE 2022\footnote{\url{https://ieeexplore.ieee.org/xpl/conhome/9793835/proceeding?isnumber=9793541}}.
For each publication, we collected the link where the code of the publication is available. However, many of the publications do not have this information: for ICSE, only 32 papers have a link and for VDLB, only 19. From the ones with code, we selected five publications by randomly generating five numbers for each conference, each number representing one publication. For ICSE, we generated five numbers between 1 and 32 and achieved [1,9,12,13,23]; for VDLB, we generated five numbers between 1 and 19 and achieved [5,7,11,15,17].

Moreover, we tried to find computational experiments in other areas through the Climate Change\footnote{\url{https://www.springer.com/journal/10584/}} and Nature Climate Change\footnote{\url{https://www.nature.com/nclimate/}} journals, but the papers would not include information about the experiment setup, without which we could not reproduce the experiment. Although this was not an extensive study, it is yet another evidence that reproducibility is not yet fully tackled by many researchers.
Regarding the query of medical, artificial intelligence and climate experiments, we only considered the first 100 results of each one. %, and we intended to randomly select five repositories. 
Once again, we randomly generated five numbers between 1 and 100 two times, five representing the medical repositories, five representing the artificial intelligence repositories and five representing the climate change repositories. We achieved the numbers [25, 38, 44, 58, 69] for medical repositories, the numbers [10, 24, 41, 64, 99] for artificial intelligence repositories and the numbers [14, 21, 31, 59, 78] for climate change repositories.

Furthermore, we have widened the range of experiences, and to do this, we have made two searches in the Zenodo repository. We searched by ``Medical''\footnote{\url{https://zenodo.org/search?page=4\&size=20\&q=medical\&type=software}}, ``Artificial Intelligence''\footnote{\url{https://zenodo.org/search?page=1\&size=20\&q=Artificial\%20Intelligence\&type=software}} and ``Climate Change''\footnote{\url{https://zenodo.org/search?page=1\&size=20\&q=climate\%20change\&type=software}} expressions, and we only considered software repositories.

We chose computer science conferences and medical, artificial intelligence and climate change experiments in the expectation that these papers would pose more complex challenges and a variety of experiences from various domains.%when compared to other research areas where one would expect the experiments would be simpler

In addition, we performed all the experiments available in all the papers related to our approach (see \cref{sec:related}).

%selected five articles for each conference to maximise the representativeness of the experiments (\eg, in terms of PLs used).

% ocupa demasiado espaço
%In~\cref{table:links-available}, we can see the number of papers analysed in each conference and the number of papers that have available the link where the code of the publication is available.

%\begin{table}[!htb]
%\caption{Characterizations of the publications analysed}
 % \centering
  %  \begin{tabular}{m{1.5cm}>{\centering}m{2.2cm}>{\centering}m{2.3cm}>{\centering\arraybackslash}m{0.7cm}}
   % \toprule
    %    Conference & Link available & Link not available & Total \\
     %   \midrule
    %VLDB 2021 & 19    & 11    & 30 \\
    %ICSE 2022 & 32    & 18    & 50 \\
    %\bottomrule
    %\end{tabular}
  %\label{table:links-available}
%\end{table}

%We chose a representative size of the experiments of the VLDB 2021 to test our approach. 

We then proceeded to collect all the information relevant to reproduce the experiment, including the PL and version used, dependencies necessary to be installed, if a database was necessary, and which commands were executed to get the expected results. 
In \cref{tab:experiments}, we present the relevant information related to the set of experiments chosen from VLDB (experiments E1--E5), from ICSE (E6--E10),  for climate change (E11--E15), for medical (E15--E20), for artificial intelligence (E21--E25) and from related work (E26--E31) (the entries in \textit{Repository link} are clickable in the PDF version of the paper and lead to the repository available in the paper). Furthermore, we added a column called ``Reproduced'' in each table, which represents if we achieved reproducibility using our approach (more details in \cref{sec:discussion}).

\begin{table*}[!htb]
    \centering
  \caption{Sample of reproducible experiments (NA stands for Not Applicable)}
    \begin{tabular}{l>{\centering}m{1.5cm}>{\centering}m{1.8cm}>{\centering}m{1.5cm}>{\centering}m{1.8cm}>{\centering}m{1.2cm}>
    {\centering}m{1cm}>{\centering\arraybackslash}m{1cm}}
    %\toprule
    \toprule
    Publication & PLs & Database & Repository link & Dependencies description  & How to execute & Repro- duced &  Identifier \\
    \midrule
    \myciteauthoryear{Adnan2022} & Unix Shell/Python & \cross & \href{https://github.com/STAR-Laboratory/Accelerating-RecSys-Training}{\underline{Link}} & \mycheckmark   & \mycheckmark & \mycheckmark & E1 \\
    \myciteauthoryear{Bai2022} & Python & \cross & \href{https://github.com/jiyangbai/TaGSim}{\underline{Link}} & \cross & \mycheckmark & \mycheckmark & E2 \\
    \myciteauthoryear{Chauhan2022} & C++/Python/ Unix Shell & \cross & \href{https://github.com/idea-iitd/RQuBE}{\underline{Link}} & \cross & \mycheckmark & \mycheckmark& E3 \\
    \myciteauthoryear{Sun2022}  & Python & PostgreSQL & \href{https://github.com/jt-zhang/CardinalityEstimationTestbed}{\underline{Link}} & \cross & \mycheckmark & \mycheckmark & E4 \\
    \myciteauthoryear{Zhou2022} & C++ & \cross & \href{https://github.com/AlexanderTZhou/IUBFC}{\underline{Link}} & \mycheckmark & \mycheckmark & \mycheckmark & E5 \\
   
   \midrule
    \myciteauthoryear{Chen2022}  & Python & \cross    & \href{https://github.com/OpsPAI/ADSketch}{\underline{Link}} & \mycheckmark & \mycheckmark & \mycheckmark& E6 \\
    \myciteauthoryear{Kukucka2022} & Java/Python/ Unix Shell & \cross & \href{https://github.com/neu-se/CONFETTI}{\underline{Link}} & \mycheckmark & \mycheckmark & \mycheckmark& E7 \\
    \myciteauthoryear{Nguyen2022} & Python/Java & \cross & \href{https://github.com/hub-se/BeDivFuzz}{\underline{Link}} & \mycheckmark & \mycheckmark & \cross & E8 \\
    \myciteauthoryear{Xie2022}  & Python & \cross    & \href{https://github.com/ICSE2022FL/ICSE2022FLCode}{\underline{Link}} & \mycheckmark & \mycheckmark  & \mycheckmark& E9 \\
    \myciteauthoryear{Zhang2022} & Java & MySQL & \href{https://buildsheriff.github.io/}{\underline{Link}} & \mycheckmark & \cross & \cross& E10 \\
    
    \midrule
    \myciteauthoryear{edwards2022}  & R & \cross & \href{https://zenodo.org/record/5967578#.ZEwQKnbMJD8}{\underline{Link}} & \mycheckmark & 
    Package & NA & E11 \\
    \myciteauthoryear{Hemes2023}  & R & \cross & \href{https://zenodo.org/record/7651416#.ZEwPxXbMJD8}{\underline{Link}} & \mycheckmark & \cross  & \mycheckmark & E12 \\
    \myciteauthoryear{Jehn2021} & Python & \cross & \href{https://zenodo.org/record/4588383#.ZEwRHnbMJD8}{\underline{Link}} & \cross & \cross & \mycheckmark & E13 \\
    \myciteauthoryear{Lin2023} & Python & \cross & \href{https://zenodo.org/record/7613549#.ZEwQSnbMJD8}{\underline{Link}} & \cross & \mycheckmark & \mycheckmark & E14 \\
    \myciteauthoryear{Qasmi2021} & R & \cross & \href{https://zenodo.org/record/5233947#.ZFPecHbMJD8}{\underline{Link}} & \mycheckmark & Package & NA & E15 \\

    \midrule
    \myciteauthoryear{Frie2021} & Python & \cross & \href{https://zenodo.org/record/4497214#.ZFQy33bMJD-}{\underline{Link}} & \cross & \mycheckmark & \mycheckmark & E16 \\
    \myciteauthoryear{Tianyu2021}  & Python & \cross & \href{https://zenodo.org/record/4926118#.ZEFWgnbMJD8}{\underline{Link}} & \mycheckmark & \mycheckmark & \cross& E17 \\

    \myciteauthoryear{Hoyt2018} & Python & \cross & \href{https://zenodo.org/record/1488650#.ZFQw4nbMJD_}{\underline{Link}} & \mycheckmark & \mycheckmark & \mycheckmark & E18 \\
    \myciteauthoryear{Kailas2022}  & Python/Unix Shell & \cross & \href{https://zenodo.org/record/6617957#.ZEFTEXbMJD8}{\underline{Link}} & \mycheckmark & \mycheckmark  & \mycheckmark& E19 \\

    \myciteauthoryear{Trueto2021} & Python & \cross & \href{https://zenodo.org/record/4453803#.ZFQxW3bMJD-}{\underline{Link}} & \mycheckmark & \mycheckmark & \mycheckmark& E20 \\
  \midrule
    \myciteauthoryear{Prezja2013} & Python & \cross & \href{https://zenodo.org/record/7867100}{\underline{Link}} & \mycheckmark & Package & NA & E21 \\
    \myciteauthoryear{Santucci2023} & Python & \cross & \href{https://zenodo.org/record/7595510}{\underline{Link}} & \cross & \cross & \mycheckmark & E22 \\
    \myciteauthoryear{Erickson2019} & Python/Unix Shell/ C++ & \cross & \href{https://zenodo.org/record/3568468}{\underline{Link}} & \halfcheckmark &  \mycheckmark & \cross & E23 \\
    \myciteauthoryear{Coupette2022} & Python /Jupyter Notebook & \cross & \href{https://zenodo.org/record/6468193}{\underline{Link}} & \mycheckmark & \cross & \mycheckmark & E24 \\
    \myciteauthoryear{Pala2023} & Python & \cross & \href{https://zenodo.org/record/7585845}{\underline{Link}} & \mycheckmark & \cross & \mycheckmark & E25 \\

    \midrule
     \myciteauthoryear{e26}  & Python/ C/Unix Shell & \cross & \href{https://github.com/cooperative-computing-lab/cctools}{\underline{Link}} & \mycheckmark & Package &  NA & E26  \\
    \myciteauthoryear{Haiyan2015}  & -- &  -- &  No info & -- & -- & NA & E27 \\
    \myciteauthoryear{Pham2015}  & -- &  -- &  No info & -- & -- & NA &  E28 \\
    \myciteauthoryear{e29}  & R & \cross & \href{https://chicago.github.io/food-inspections-evaluation/}{\underline{Link}} & \mycheckmark & \mycheckmark & \mycheckmark & E29 \\
   \myciteauthoryear{e30}    & Python/Unix Shell & \cross & \href{https://github.com/uva-hydroinformatics/VIC_Pre-Processing_Rules}{\underline{Link}} & \cross & \cross  & \cross& E30 \\
   \myciteauthoryear{e31}   & Python & SQlite & \href{https://bitbucket.org/TonHai/iqe/src/master/}{\underline{Link}} & \cross & \cross & \mycheckmark & E31 \\
    \bottomrule
    %\bottomrule
    \end{tabular}%
  \label{tab:experiments}%
\end{table*}%

%Some experiments, which were not chosen, were not possible to reproduce due to the need for computing power or because it was necessary to install specific frameworks and compilers, so we did not consider those cases.

%In our approach, we support all the Python, C++, Java, Shell and Jupyter Notebook versions available in the Docker Hub repository. 

%Related to the compilers, the compiler used to build Java projects is the Maven. Ant and Gradle were not considered in this approach. The compiler used to compile C++ projects is the GCC and G++, one more time, we can support all the versions available in the Docker Hub repository.

Our own experiment environment is a personal computer with a 6-core CPU, and 16GB RAM, running Ubuntu 20.04.

\subsection{Execution}
Our approach is the backend of the platform, not providing a user interface, so we need to use its API on this execution.
For each of the projects, we started by creating and populating with the corresponding code the projects (\cref{sec:ex_create}). We then build the necessary environment (\cref{sec:ex_env}) and execute the experiments (\cref{subsection:runExperiment}). Finally, we create a reproducibility package (\cref{subsection:reproducible_package}).

\subsubsection{Create a project and upload its files}\label{sec:ex_create}
During this task, we used the ``Data Management Component'' described in \cref{sec:data}.
The first step is to create a project and fill in some important fields, such as the project name and the project description. The response is the unique identifier of the project created. 
During the following tasks, we will need the returned unique identifier of the project.

The second step is to upload files to the project. We can upload a zip file or use a remote location (\git or DOI). %In both cases, the response is a message and a status code.
At the end of this task, a new project is created in the platform.

\subsubsection{Build the environment for each experiment}\label{sec:ex_env}
The main goal of this task is to recreate the same computational environment of an experience.
During this task, we used the ``Computational Environment Component'' described in \cref{sec:environment}.
We will exemplify the building process of two environments for two distinct experiments(E3 and E8).

The order in which the experiments are performed is irrelevant, and we could choose another experiment, but we consider that the first experiment we want to reproduce is the E3, and the second one is the E8.
E3 uses C++, Python, Jupyter Notebook, and Unix Shell as the PLs. Furthermore, this experiment has only one Jupyter Notebook file. Python and Unix Shell files were not used in this execution, so we do not consider these PLs.
The documentation of E3 indicates that to build the experiment, it is necessary to run the commands \verb|g++ -O3 ./src/bbfs\_node.cpp -o out| and \verb|g++ -O3 ./src/bbfs\_edge.cpp -o out|.
For this task, we make use of the ``Build a docker image'' functionality of the ``Computational Environment Component'' (\cref{sec:environment}). Our approach infers the PLs (C++) and inserts the commands to build the project.
Listing~\ref{lst:requestData1} represents the request data after the user fills in all the information.

\begin{lstlisting}[language=json,firstnumber=1, caption={Request data of E3},captionpos=b, label={lst:requestData1},float ]
{"languages": ["C++"],
"languagesVersion":{"C++":"gcc:8"},
"commandsToAdd":[
    "g++ -O3 ./src/bbfs_node.cpp -o out", 
    "g++ -O3 /src/bbfs_edge.cpp -o out"]}
\end{lstlisting}

After receiving this request, the server generates a dockerfile, builds a docker image and returns an identifier (ID) that represents that docker image.
In Listing~\ref{lst:dockerfile4}, we can see the dockerfile generated for the E3. The directory ``files'' contain all the files of the experiment. 

\begin{lstlisting}[language=json,firstnumber=1, caption={Dockerfile of E3},captionpos=b, label={lst:dockerfile4},float ]
FROM ubuntu:20.04
RUN  apt update &&  apt upgrade -y
RUN apt install -y gcc-8 make g++
RUN update-alternatives --install /usr/bin/gcc gcc /usr/bin/gcc-8 2000
WORKDIR /files
COPY ./files .
RUN g++ -O3 ./src/bbfs_node.cpp -o out
RUN g++ -O3 ./src/bbfs_edge.cpp -o out
\end{lstlisting}

%The second experiment, which we intend to reproduce, is the E8.
%
E8 is a more complex experiment than E3 as it needs three PLs to execute (Python, Java and Unix Shell).
The authors of the experiment mentioned that the prerequisites are ``Java 8-11, Apache Maven and Python 3''.
Furthermore, all the Python dependencies that are necessary to install are in a file in the repository.
To conclude, it is necessary to execute the following commands \verb|cd RLCheck/jqf/|, \verb|mvn package|, and \verb|cd ../..|. Our approach needs to replace \verb|cd| command to \verb|WORKDIR| in the generations of the dockerfile because the \verb|cd| command has no effect after the execution of that command, and \verb|WORKDIR| command changes the working directory for future commands.
Thus, the request type to generate a dockerfile is the same as E3, but the content of the request data is different and is represented in Listing~\ref{lst:requestData2}.

\begin{lstlisting}[language=json,firstnumber=1, caption={Request data of E8},captionpos=b, label={lst:requestData2}, float ]
{"languages": ["python", "shell", "java"],
"languagesVersion":{"java":"openjdk-11", "python":"python:3.8"},
"commandsToAdd":["cd RLCheck/jqf/","mvn package","cd ../.."],
"hasRequirementsFile": "true"}
\end{lstlisting}

In Listing~\ref{lst:dockerfile8}, we can see the dockerfile generated for E8.

\begin{lstlisting}[language=json,firstnumber=1, caption={Dockerfile of E8},captionpos=b, label={lst:dockerfile8}, float ]
FROM ubuntu:20.04
RUN  apt update &&  apt upgrade -y
RUN apt install -y python3.8 python3-pip
RUN update-alternatives --install /usr/local/bin/python python /usr/bin/python3.8 2000
RUN update-alternatives --install /usr/bin/pip pip /usr/bin/pip3 2000
RUN apt install -y openjdk-11-jdk openjdk-11-jre maven
WORKDIR /files
COPY ./files .
RUN mvn package
WORKDIR RLCheck/jqf/
RUN mvn package
WORKDIR ../..
RUN pip install -r requirements.txt
\end{lstlisting}

\subsubsection{Run the experiments}\label{subsection:runExperiment}
After we have a docker image created, we only need to execute the experiment. For that to be possible, we need to insert the Id of the created docker image in the previous task and which command we intend to run.
To do so, we make use of the ``run an experiment'' functionality of the ``Code Execution Component'' (\cref{sec:code}).

Listing~\ref{lst:requestDataRunExperiment} presents the request data to execute E3.

\begin{lstlisting}[language=json,firstnumber=1, caption={Request data for the execution of E3},captionpos=b, label={lst:requestDataRunExperiment}, float ]
{"command": "./out freebase/edges.txt freebase/labels.txt 1234 5678 4 0.3 2000 0.8 1 10 0.01 2",
"tagId": 100}
\end{lstlisting}

\subsubsection{Build the reproducible package}\label{subsection:reproducible_package}
Finally, we created a reproducibility package for each of the experiments, which we make publicly available\footnote{\url{https://zenodo.org/record/8050716}}.

We reuse the container image generated in the building of the computational environment task and reuse all the information provided on the execution of the experiment (commands) to encapsulate all that information in a folder along with the scripts needed to run the project. We built a reproducible package that can be run on Windows or Unix operating systems where the user just needs to have Docker installed and running. By clicking in ``runExperiment'' file, the experiment is executed automatically.

\section{Discussion}\label{sec:discussion}
We chose 31 different experiments from published work in two research communities (ICSE and VLDB), in three research areas (Medical, Artificial Intelligence and Climate Change) and experiments executed in related work approaches. 

Although 31 may not seem a significant number, making each one reproducible was quite challenging due to the diversity of experiments considered, encompassing different domains, and the variety of computational procedures involved.
During the configuration of the computational environment, we needed to infer the information from the repository (\eg, PLs and dependencies used ) to try to fill in the parameters that were necessary to insert to build the same computational environment. This information was only sometimes clear. 
Our approach can infer the Python and R dependencies needed when the requirements file is missing. However, deducing the dependencies for other PLs is very difficult, so that information should be part of the project description. Although our backend already includes such a feature, we must design an easy way for researchers to have that information in the frontend.

We analyzed the articles on all the related work approaches. However, most methods do not have a validation section (Open Lab~\cite{Shevchenko2022}, Encapsulator~\cite{Pasquier2018}, VisTrails~\cite{Freire2012}, RenkuLab~\cite{krieger2021repeatable}, Whole Tale~\cite{Brinckman2019}, ReproZip~\cite{Chirigati2013,Chirigati2016,Chirigati2016_colaborative}, RunMyCode~\cite{Stodden2012} and Binder~\cite{Matthias2018binder}). It was thus not possible to compare our approach with those.
An exception is CARE~\cite{Janin2014}. The authors selected seven significant applications (\eg, VLC, Firefox). However, they did not reproduce experiments but rather executed commands. Thus, there is no experiment to be reproduced. Nevertheless, our platform supports the PLs of that commands. 

The authors of E26 and E27 presented two case studies of complex, high-energy physics applications and showed how their framework could be useful for sandboxing, preserving, and distributing applications~\cite{Haiyan2015}. Although the authors make available the case study link \footnote{\url{https://curate.nd.edu/downloads/5712m615h8d}}, only one case study is actually available. Nerveless, it is impossible to reproduce the experiment because they only distribute a software package for enabling large-scale distributed computing, and there is no expected result, so we cannot compare our approach with this.

CDE-SP's authors described a pipeline, identified by E28,  they conducted to determine the overall performance of CDE-SP~\cite{Pham2015}. However, they do not make available the experiment code, so we cannot reproduce the same experiment.

%SciInc~\cite{Youngdahl2019}, 

SciInc's authors selected three use cases for their evaluation (E29~\cite{e29}, E30~\cite{e30}, E31~\cite{e31}), and it is possible to access the repository~\cite{Youngdahl2019}. We can install and execute E29 and E31, but it is not possible to verify the computational reproducibility because there is no information about the commands executed and the expected results. E30 does not have information about the experiment, but we tried to execute the ``Main\_Shell\_Script.scr'' file in the root of the project. However, it was impossible to reproduce that file since it tries to access information inside the \verb|cmd| folder, but there is no such folder in the repository, so not all the files from this repository have been published.

The other three experiments we could not reproduce (E8, E10, E17 and E23) have different causes. For E8, the problem was due to not finding a Java class during the execution, E10 does not exchange information between the database and the functions of the experiment, for E17, the necessary dependencies have been installed, but the data needed to run the experiment is not available. We asked permission to access the CheXpert\footnote{\url{https://stanfordmlgroup.github.io/competitions/chexpert/}} portal; however, it was impossible to find the files inside the portal.
For E23, there is no information about how to install the project, so we tried to install the project using ''python setup.py install`` command, however, during the installation, some files were missing. 

Nevertheless, from the 31 chosen experiments, four of them are packages, and two do not have available the source code, so only 25 experiments are available to be reproduced. We were able to successfully install the required dependencies, execute, and create a reproducibility package for 20 of 25 (80\%). Note that we were not even able to run E8, E10, E17, E23 and E30 on our computer (without the platform), and thus we feel confident stating that this was not an issue of our platform.

The execution of the experiments went smoothly. However, although the repositories included the command lines to execute the code, they did not have the expected results, so the verification of the reproducibility became more difficult, but we managed to compare our results with the claims of developments in the paper.

We are confident the challenges described in \cref{sec:challengesInReproducibility} (C1, C2, C3) are well addressed by our approach.

Software (C1) is the biggest problem when addressing reproducibility. Our proposal allows users to build a complete container of the project, which will make it available for reuse by other researchers. In this way, it is possible to execute the same computational environment, even when some dependencies are not publicly available anymore. This addresses principles P1 and P2.

Command (C2) implemented in our approach allows researchers to configure the execution of experiments and execute the experiment with the same parameters. The results achieved by previous executions can be used as a means of comparison to validate the computational reproducibility. Principles P3 and P4 are tackled in this part of the work. 

Data (C3) is handled in the platform by providing a mechanism to allow researchers to manage the data of the experiment. This was also our principle P5. 

Our approach is able to deal with a diversity of PLs (and their dependencies) used by
researchers. Furthermore, it supports a variety of experiments, from the execution of simple scripts to the execution of more complex experiments (\eg, involving databases or several PLs). Thus, we can say principles P6 and P7 are also addressed by our proposal.

Moreover, the reproducibility package that can be obtained from the platform can be easily shared as it corresponds to a compressed file with just two files. By executing one, with a simple double-click, the entire experiment is executed. The only requirement is that docker, which is available both for Unix and Windows systems, is installed and running. Thus, we have reduced the problem of re-execution to having to install a single software tool, the most used for container management. Thus, principles P9 and P10 are also addressed. 

Our approach is the backend of a platform, not providing a user interface. Thus, principle P8, which we consider a very important one, is not yet supported by our work.
Nevertheless, we plan to design and build such an interface in future work \cite{vlhcc2022}.

\section{Threats to Validity}\label{sec:threats}

The validation process has some threats divided into 4 sections~\cite{Cook1979, Wohlin2012} addressed next.
%``Conclusion validity''(ConcV), ``Internal validity''(IV), ``Construct validity''(ConsV) and ``External validity''(EV).  For each section, we collected some threats that we need to pay attention to in our validation process.

%For ConcV section, we choose two important threats for our validation.
\subsection*{Conclusion validity}

\paragraph{Random heterogeneity of subjects} 
 We tried to collect a variety of experiments, including experiments with only one PL and experiments with more than one, and with and without the use of databases. Moreover, we have selected experiments from different scientific fields, including climate change and medicine. Thus, our subjects are relatively heterogeneous and represent a sample of published scientific experiments.

%For IV section, we choose two important threats for our validation
\subsection*{Internal validity}

\paragraph{Instrumentation.} The computational environment is only briefly described by the authors of the experiments we used, so it is more difficult to try to infer what dependencies and which versions were used in the execution of the original experiment. Thus, there is the risk that we have executed the experiments in slightly different conditions. Nevertheless, we were able to validate the results we obtained with the ones in the papers, and thus this risk is quite low.

%\paragraph{Fishing and the error rate.} During the development of our approach, we were looking for errors in the execution of experiments, so the portion of successfully executed experiments using other experiments can be different.

 %repetida
%\paragraph{Reliability of treatment implementation.} Researchers did not describe what dependencies and which versions were used in the execution of the experiments, so our computational environment built for the execution can be different.

\subsection*{Construct validity}
%For ConsV section, we choose one important threat for our validation. 

% é igual à anterior
%\paragraph{Inadequate preoperational explication of constructs.} Such as we mentioned in previous threats, the description of the computational environment used by the authors is not clear. Furthermore, we do not know, in detail, the experiment executed by the authors.

\paragraph{Restricted generalizability across constructs.} Our approach improves computational reproducibility, however, for some experiments it is necessary to have more storage and computational power. %$ to execute the same experiment but using our approach. 
Thus, we were not able to validate our approach with such demanding requirements. Nevertheless, most of the experiments we analyzed during this work did not have such restrictions and thus could potentially be executed within our platform.

\subsection*{External validity}
%For EV section, we choose two important threats for our validation. 

\paragraph{Interaction of selection and treatment.} We collected a set of experiments from different computer science areas, however, we do not have experiments from domains other than computer science. 
Nevertheless, from experience, we know that at least in some areas it is not common to report the computational setting and thus it was not possible to find experiments from other domains. Moreover, the experiments we used are quite complex (\eg, in terms of (the number of) PLs used, databases, etc.) and thus we are confident that our platform can cope with experiments from other domains.
% Moroever, there are a variety of experiments that we do not reproduce (\eg, high computational demanding experiments).

%\paragraph{Interaction of setting and treatment.} acho que nao se aplica

\section{Related Work}\label{sec:related}
%\citestyle{acmauthoryear}
    There have been some tools proposed to aid researchers in making their computational experiments reproducible. We discuss them now based on their capabilities.
%(Open Lab~\cite{Shevchenko2022}, 
%SciInc~\cite{Youngdahl2019}, CDE-SP~\cite{Pham2015}, Encapsulator~\cite{Pasquier2018}, A Framework~\cite{Haiyan2015}, CARE~\cite{Janin2014}, VisTrails~\cite{Freire2012}, 
%Binder~\cite{Matthias2018binder}, RenkuLab~\cite{krieger2021repeatable}, 
%Whole Tale~\cite{Brinckman2019}, ReproZip~\cite{Chirigati2013,Chirigati2016,Chirigati2016_colaborative},
%RunMyCode~\cite{Stodden2012}
%). This section overviews the reproducibility challenges of various research areas and which tools or platforms can overtake these challenges.

\subsection{Supported PLs}
Researchers perform experiments using a variety of PLs, so it is essential that the platform or the tool used can support such experiments. Some related works such as Encapsulator~\cite{Pasquier2018},  Binder~\cite{Matthias2018binder}, RenkuLab~\cite{krieger2021repeatable}, Whole Tale~\cite{Brinckman2019}, Open Lab~\cite{Shevchenko2022}, VisTrails~\cite{Freire2012}, RunMyCode~\cite{Stodden2012} or the work by Haiyan~\cite{Haiyan2015} support only a limited set of PLs and are not extendable, so it is impossible to support new PLs. However, CDE-SP~\cite{Pham2015}, CARE~\cite{Janin2014}, SciInc~\cite{Youngdahl2019}, and ReproZip~\cite{Chirigati2013,Chirigati2016,Chirigati2016_colaborative} support an unlimited set of PLs.

\subsection{Automatic installation of dependencies}
We see in \cref{sec:challengesInReproducibility}, that one of the main challenges in computational reproducibility is to install the required dependencies and library correctly. In this way, it is crucial that the platform can infer and install all the necessary dependencies and libraries to execute the code in the same way as the owner of the experiment.

ReproZip~\cite{Chirigati2013, Chirigati2016, Chirigati2016_colaborative}, VisTrails~\cite{Freire2012}, CDE-SP~\cite{Pham2015}, CARE~\cite{Janin2014}, SciInc~\cite{Youngdahl2019}, and the work by Haiyan~\cite{Haiyan2015} can automatically track and identify all the required dependencies, binary files, code, data, and files needed to execute the code. Furthermore, Encapsulator~\cite{Pasquier2018} employs observed provenance capture to gather detailed provenance information and expose the underlying mechanisms of an analysis script. This process involves capturing fine-grained provenance data.

Binder~\cite{Matthias2018binder}, RenkuLab~\cite{krieger2021repeatable} and 
Whole Tale~\cite{Brinckman2019} are limited approaches because the researcher needs to install all the required dependencies in the platform.  

\subsection{Supported databases}
In general, computational experiments do not have databases integrated. However, complex experiments that involve databases are harder to reproduce, because reproducible tools do not support these experiments.
ReproZip~\cite{Chirigati2013, Chirigati2016, Chirigati2016_colaborative} is the only approach that supports experiments that have integrated databases. 

%\subsection{Reuse samples from datasets}

\subsection{Configure the execution of the experiments}
Making available all the information to allow computational reproducibility is essential to reproduce the same results; such information includes the values of the parameters used in the execution of the experiment.

One way to easily make this information available is to be possible to share the parameter spaces explored. Open Lab~\cite{Shevchenko2022}, VisTrails~\cite{Freire2012}, Whole Tale~\cite{Brinckman2019}, ReproZip~\cite{Chirigati2013, Chirigati2016, Chirigati2016_colaborative}, RunMyCode~\cite{Stodden2012} are approaches that make it possible because the owner of the experiment can share the same parameter spaces explored with other researchers and making the process of the configuration of the experiment easier in the reproduction.
%\subsection{Guided assistant}

\subsection{Reproducible package}
Docker offers greater reliability and flexibility, as well as being lightweight and capable of maintaining its own data and memory~\cite{boettiger2015, Dijiang2018}. By utilizing container images, Docker enables faster start-up times and efficient utilization of resources such as RAM and disk. This is due to the shared files and layered file systems that are inherent to containerization. By using this technique, it is possible to share a capsule with all the information needed to execute the experiment in the same way without the need to use another platform besides Docker. 

Encapsulator \cite{Pasquier2018}, SciInc \cite{Youngdahl2019}, ReproZip \cite{Chirigati2013, Chirigati2016, Chirigati2016_colaborative}, CARE \cite{Janin2014},
VisTrails~\cite{Freire2012}, CDE-SP~\cite{Pham2015}, and the work by Haiyan
\cite{Haiyan2015} are approaches that automatically package everything required to execute the same experiment in another machine without the use of a specific platform.

Binder~\cite{Matthias2018binder}, RenkuLab~\cite{krieger2021repeatable} and  Whole Tale~\cite{Brinckman2019} create a Docker image of the repository, but one needs the original platform to re-execute the code. Open Lab~\cite{Shevchenko2022}, RunMyCode~\cite{Stodden2012} are approaches that, such as the previous one, it is necessary to re-execute the code on the platform, however, they do not use virtualization. 

\subsection{Replicability of results}
Replicability and reproducibility are two concepts that produce the same result, but using different methods, code or data.
In all the related work approaches is not possible to mark or associate a dataset to each execution, making the comprehension of the dataset used harder in future executions. In this way, all the related work approaches do not validate the replicability of results.

\subsection{Validation of results}
Validating the results stated in publications is hard work. Researchers have to run the experiment, read the results section, and compare the results obtained with the results stated in the publication. It was advantageous for the researchers to enter what the expected result was in the run and in each run this comparison was made in order to prove the reproducibility of results. However, the approaches in the related papers do not support this functionality.

\subsection{Supported operating system}
Researchers build their experiments in their favorite Operating Systems (OS), so it should not be a limitation when we intend to reproduce an experiment.
Binder~\cite{Matthias2018binder}, RenkuLab~\cite{krieger2021repeatable}, Whole Tale~\cite{Brinckman2019}, Open Lab~\cite{Shevchenko2022}, VisTrails~\cite{Freire2012}, RunMyCode~\cite{Stodden2012} and Encapsulator~\cite{Pasquier2018} are approaches that support all the Operating Systems, so it is not a limitation. However, Encapsulator~\cite{Pasquier2018}, 
SciInc~\cite{Youngdahl2019}, 
 CARE~\cite{Janin2014}, 
ReproZip~\cite{Chirigati2013,Chirigati2016,Chirigati2016_colaborative}
and the work by \myciteauthoryear{Haiyan2015} are approaches limited to the Linux environment. This might prove problematic for many scientists who do not use Linux. Furthermore, CDE-SP~\cite{Pham2015} only supports Linux and Mac OS environments.

We conclude that the main platforms and approaches for sharing and computing research data are focused on specific domains and thus they are not able to reproduce complex experiments (using several environments, databases, etc.) or are not user-friendly platforms.
%, so we propose for our approach to try to solve these limitations.
In \cref{tab:toolComparisonwithGUI} and \cref{tab:toolComparisonwitoutGUI} we synthesize the characteristics of the methods and tools that we have just described.

%\todo{nestas tabelas temos de colocar a nossa proposta tb para se comparar melhor}

\begin{table*}[!htb]
\caption{Comparison of characteristics of reproducibility tools with Graphical User Interface (GUI)}
  \centering
    \begin{tabular}{m{5.3cm} >{\centering}m{.85cm} >{\centering}m{.85cm}>{\centering}m{.7cm}>{\centering} m{1cm}>{\centering}m{.7cm}>{\centering}m{.8cm}>{\centering\arraybackslash}m{1cm} c}
    \toprule
          &  Binder & Renku-Lab & Whole Tale & Repro- Zip  & Open Lab Me & Vis- Trails & RunMy- Code & Our\\
     \midrule
    Unlimited PLs  &\cross  & \cross & \cross & \mycheckmark& \cross & \cross& \cross & \mycheckmark\\
    Researcher-friendly platform  & \mycheckmark  & \cross & \mycheckmark & \cross & \mycheckmark & \mycheckmark&\mycheckmark& \cross\\
    Data management & \mycheckmark & \mycheckmark & \mycheckmark & \halfcheckmark & \mycheckmark & \mycheckmark & \mycheckmark & \mycheckmark\\
    %Metadata cataloging & \cross  & \cross & \halfcheckmark & \cross & \cross & \cross &\cross& \cross\\
    %Metadata advanced search  & \cross  & \cross & \cross & \cross & \cross & \cross&\cross& \cross\\
    %Reuse samples from datasets  & \cross  & \cross & \halfcheckmark & \halfcheckmark & \cross & \cross&\cross& \cross\\
    Supports databases  &\cross  & \cross & \cross & \mycheckmark& \cross & \cross&\cross& \mycheckmark\\
     Supports experiments from different fields & \halfcheckmark & \halfcheckmark & \halfcheckmark & \halfcheckmark & \halfcheckmark & \halfcheckmark & \halfcheckmark & \mycheckmark  \\

     Automatically infers dependencies  & \cross  & \cross & \cross & \mycheckmark&  \cross & \mycheckmark & \cross & \halfcheckmark\\

     Automatically infers PLs &\cross  & \cross & \cross & \mycheckmark& \cross & \cross& \cross & \halfcheckmark\\
    
    Automatically installs necessary \qquad dependencies  & \halfcheckmark  & \halfcheckmark & \halfcheckmark & \mycheckmark& \cross & \mycheckmark&\cross& \mycheckmark\\
    
    Configures execution of experiments  &\cross  & \cross & \mycheckmark & \mycheckmark& \mycheckmark & \mycheckmark&\mycheckmark& \mycheckmark\\
    %Guided assistant  &\cross  & \cross & \cross & \cross& \cross & \cross& \cross& NA\\
     %\gls{GUI} & \mycheckmark &\mycheckmark  & \mycheckmark & \mycheckmark & Repro-Unzip plugin & \mycheckmark & \mycheckmark&\mycheckmark\\
     Creation of reproducibility package & \cross & \cross & \cross & \mycheckmark & \cross& \mycheckmark & \cross & \mycheckmark\\
     Reproducibility package executed anywhere & \cross & \cross & \cross & \mycheckmark  & \cross & \mycheckmark  & \cross & \mycheckmark\\
     Validation of results & \cross & \cross & \cross & \cross & \cross & \cross& \cross& \mycheckmark\\
     Supported operating system  & All & All & All  & Linux & All & All & All& All \\
    \bottomrule
    \end{tabular}%
  \label{tab:toolComparisonwithGUI}%
\end{table*}%

\begin{table*}[!htbp]
\caption{Comparison of characteristics of reproducibility tools without GUI (only command line interface)}
  \centering
    \begin{tabular}{m{5.3cm}cccccc}
    \toprule
          & CDE-SP & Encapsulator & \cite{Haiyan2015} & CARE & SciInc & Our \\
     \midrule
    Unlimited PLs & \mycheckmark & \cross & \cross  &\mycheckmark  & \mycheckmark & \mycheckmark  \\
    Data management  & \cross & \cross & \cross &\cross & \cross & \mycheckmark \\
    %Metadata cataloging & \cross& \cross  & \cross  & \cross & \cross  & \cross\\
    %Metadata advanced search & \cross&  \cross  &\cross  & \cross & \cross  & \cross \\
    %Reuse samples from datasets & \cross & \cross  & \cross & \cross & \cross & \cross   \\
     Supports experiments from different fields & \halfcheckmark & \halfcheckmark & \halfcheckmark & \halfcheckmark & \halfcheckmark & \mycheckmark    \\
    
    Supports databases & \cross &\cross  & \cross & \cross & \cross & \mycheckmark \\
    Automatically infers dependencies & \mycheckmark & \mycheckmark  & \mycheckmark & \mycheckmark & \mycheckmark & \halfcheckmark\\

    Automatically infers PLs & \mycheckmark & \cross & \cross  &\mycheckmark  & \mycheckmark & \halfcheckmark\\
    
    Automatically installs  dependencies & \mycheckmark & \mycheckmark  & \mycheckmark & \mycheckmark & \mycheckmark & \mycheckmark\\
    Configures execution of experiments & \cross &\cross  & \cross & \cross & \cross & \mycheckmark \\
    %Guided assistant & \cross &\cross  & \cross & \cross & \cross & NA\\
     Creation of reproducibility package & \mycheckmark & \mycheckmark & \mycheckmark &\mycheckmark & \mycheckmark& \mycheckmark\\
     Reproducibility package executed anywhere & \mycheckmark & \mycheckmark & \mycheckmark &\mycheckmark & \mycheckmark& \mycheckmark\\
     Validation of results & \cross & \cross & \cross & \cross & \cross & \mycheckmark\\
    Supported operating system & Linux/Mac OS & All & Linux &  Linux  &  Linux & All \\
    \bottomrule
    \end{tabular}\\
  \label{tab:toolComparisonwitoutGUI}%
\end{table*}%

\section{Conclusion}\label{sec:conclusion}
Our approach is designed to support researchers in ensuring their research is reproducible and transparent during their study and in preparation for publication. One of the biggest obstacles to achieve this is creating the same computational environment, configuring it correctly, reproducing the experiment, and verifying its computational reproducibility.

We propose a solution to address these challenges and assist researchers in developing the experiment and preparing a reproducibility package to accompany their paper. This will enable reviewers and others to rerun the experiments. Our evaluation of the approach revealed that we successfully recreated and reproduced 20 out of 25 experiments, demonstrating that our methodology and backend platform can easily recreate, reproduce, and share experiments.

Currently, we are developing a frontend interface to enable researchers to create reproducible packages with ease \cite{vlhcc2022}. 
Moreover, we plan to further study how to make the sharing of data among experiments easier for researchers, but also for the community in general \cite{LCosta2019}.
%With the validation of our backend platform already complete, our next focus is to validate the usability of the user interface.

%----------------------------------------------------------------------------------------
%	 REFERENCES
%----------------------------------------------------------------------------------------

%\printbibliography % Output the bibliography

\bibliographystyle{plainnat}

\bibliography{bibliography.bib} % BibLaTeX bibliography file

\begin{thebibliography}{65}
\providecommand{\natexlab}[1]{#1}
\providecommand{\url}[1]{\texttt{#1}}
\expandafter\ifx\csname urlstyle\endcsname\relax
  \providecommand{\doi}[1]{doi: #1}\else
  \providecommand{\doi}{doi: \begingroup \urlstyle{rm}\Url}\fi

\bibitem[Adnan et~al.(2022)Adnan, Maboud, Mahajan, and Nair]{Adnan2022}
Muhammad Adnan, Yassaman~Ebrahimzadeh Maboud, Divya Mahajan, and Prashant~J.
  Nair.
\newblock Accelerating recommendation system training by leveraging popular
  choices.
\newblock \emph{Proc. VLDB Endow.}, 15\penalty0 (1):\penalty0 127–140, jan
  2022.
\newblock ISSN 2150-8097.
\newblock \doi{10.14778/3485450.3485462}.
\newblock URL \url{https://doi.org/10.14778/3485450.3485462}.

\bibitem[Bai and Zhao(2022)]{Bai2022}
Jiyang Bai and Peixiang Zhao.
\newblock Tagsim: Type-aware graph similarity learning and computation.
\newblock \emph{Proc. VLDB Endow.}, 15\penalty0 (2):\penalty0 335–347, feb
  2022.
\newblock ISSN 2150-8097.
\newblock \doi{10.14778/3489496.3489513}.
\newblock URL \url{https://doi.org/10.14778/3489496.3489513}.

\bibitem[Boettiger(2015)]{boettiger2015}
Carl Boettiger.
\newblock An introduction to docker for reproducible research.
\newblock \emph{SIGOPS Oper. Syst. Rev.}, 49\penalty0 (1):\penalty0 71–79,
  jan 2015.
\newblock ISSN 0163-5980.
\newblock \doi{10.1145/2723872.2723882}.
\newblock URL \url{https://doi.org/10.1145/2723872.2723882}.

\bibitem[Brinckman et~al.(2019)Brinckman, Chard, Gaffney, Hategan, Jones,
  Kowalik, Kulasekaran, Ludäscher, Mecum, Nabrzyski, Stodden, Taylor, Turk,
  and Turner]{Brinckman2019}
Adam Brinckman, Kyle Chard, Niall Gaffney, Mihael Hategan, Matthew~B. Jones,
  Kacper Kowalik, Sivakumar Kulasekaran, Bertram Ludäscher, Bryce~D. Mecum,
  Jarek Nabrzyski, Victoria Stodden, Ian~J. Taylor, Matthew~J. Turk, and
  Kandace Turner.
\newblock Computing environments for reproducibility: Capturing the “whole
  tale”.
\newblock \emph{Future Generation Computer Systems}, 94:\penalty0 854--867,
  2019.
\newblock ISSN 0167-739X.
\newblock \doi{10.1016/j.future.2017.12.029}.
\newblock URL \url{https://doi.org/10.1016/j.future.2017.12.029}.

\bibitem[Brunsdon and Comber(2021)]{Brunsdon2021}
Chris Brunsdon and Alexis Comber.
\newblock {Opening practice: supporting reproducibility and critical spatial
  data science}.
\newblock \emph{Journal of Geographical Systems}, 23\penalty0 (4):\penalty0
  477--496, oct 2021.
\newblock ISSN 14355949.
\newblock \doi{10.1007/s10109-020-00334-2}.

\bibitem[Bush et~al.(2020)Bush, Dutton, Evans, Loft, and Schmidt]{Bush2020}
Rosemary Bush, Andrea Dutton, Michael Evans, Rich Loft, and Gavin~A. Schmidt.
\newblock Perspectives on {Data} {Reproducibility} and {Replicability} in
  {Paleoclimate} and {Climate} {Science}.
\newblock \emph{Harvard Data Science Review}, 2\penalty0 (4), dec 16 2020.
\newblock \doi{10.1162/99608f92.00cd8f85}.
\newblock https://hdsr.mitpress.mit.edu/pub/dijwtzza.

\bibitem[Bussonnier et~al.(2018)Bussonnier, Forde, Freeman, {G}ranger, {H}ead,
  Holdgraf, {K}elley, {G}ladys Nalvarte, {O}sheroff, {P}acer, {Y}uvi {P}anda,
  {F}ernando {P}erez, {B}enjamin~{R}agan {K}elley, and
  {W}illing]{Matthias2018binder}
Matthias Bussonnier, Jessica Forde, Jeremy Freeman, Brian {G}ranger, Tim
  {H}ead, Chris Holdgraf, Kyle {K}elley, {G}ladys Nalvarte, Andrew {O}sheroff,
  {M} {P}acer, {Y}uvi {P}anda, {F}ernando {P}erez, {B}enjamin~{R}agan {K}elley,
  and Carol {W}illing.
\newblock Binder 2.0 - {R}eproducible, interactive, sharable environments for
  science at scale.
\newblock In {F}atih {A}kici, {D}avid {L}ippa, {D}illon {N}iederhut, and {M}
  Pacer, editors, \emph{Proceedings of the 17th Python in {S}cience
  {C}onference}, pages 113--120, Austin, Texas, 2018. SciPy.
\newblock \doi{10.25080/Majora-4af1f417-011}.
\newblock URL \url{https://doi.org/10.25080/Majora-4af1f417-011}.

\bibitem[Chauhan et~al.(2022)Chauhan, Jain, Ranu, Bedathur, and
  Bagchi]{Chauhan2022}
Komal Chauhan, Kartik Jain, Sayan Ranu, Srikanta Bedathur, and Amitabha Bagchi.
\newblock Answering regular path queries through exemplars, feb 2022.
\newblock ISSN 2150-8097.
\newblock URL \url{https://doi.org/10.14778/3489496.3489510}.

\bibitem[Chen et~al.(2022)Chen, Liu, Su, Zhang, Ling, Yang, and Lyu]{Chen2022}
Zhuangbin Chen, Jinyang Liu, Yuxin Su, Hongyu Zhang, Xiao Ling, Yongqiang Yang,
  and Michael~R. Lyu.
\newblock Adaptive performance anomaly detection for online service systems via
  pattern sketching.
\newblock In \emph{Proceedings of the 44th International Conference on Software
  Engineering}, ICSE '22, page 61–72, New York, NY, USA, 2022. Association
  for Computing Machinery.
\newblock ISBN 9781450392211.
\newblock \doi{10.1145/3510003.3510085}.
\newblock URL \url{https://doi.org/10.1145/3510003.3510085}.

\bibitem[Chirigati et~al.(2013)Chirigati, Shasha, and Freire]{Chirigati2013}
Fernando Chirigati, Dennis Shasha, and Juliana Freire.
\newblock Packing experiments for sharing and publication.
\newblock In \emph{Proceedings of the 2013 ACM SIGMOD International Conference
  on Management of Data}, SIGMOD '13, page 977–980, New York, NY, USA, 2013.
  Association for Computing Machinery.
\newblock ISBN 9781450320375.
\newblock \doi{10.1145/2463676.2465269}.
\newblock URL \url{https://doi.org/10.1145/2463676.2465269}.

\bibitem[Chirigati et~al.(2016{\natexlab{a}})Chirigati, Capone, Rampin, Freire,
  and Shasha]{Chirigati2016_colaborative}
Fernando Chirigati, Rebecca Capone, R\'{e}mi Rampin, Juliana Freire, and Dennis
  Shasha.
\newblock A collaborative approach to computational reproducibility.
\newblock \emph{Inf. Syst.}, 59\penalty0 (C):\penalty0 95–97, jul
  2016{\natexlab{a}}.
\newblock ISSN 0306-4379.
\newblock \doi{10.1016/j.is.2016.03.002}.
\newblock URL \url{https://doi.org/10.1016/j.is.2016.03.002}.

\bibitem[Chirigati et~al.(2016{\natexlab{b}})Chirigati, Rampin, Shasha, and
  Freire]{Chirigati2016}
Fernando Chirigati, R\'{e}mi Rampin, Dennis Shasha, and Juliana Freire.
\newblock Reprozip: Computational reproducibility with ease.
\newblock In \emph{Proceedings of the 2016 International Conference on
  Management of Data}, SIGMOD '16, page 2085–2088, New York, NY, USA,
  2016{\natexlab{b}}. Association for Computing Machinery.
\newblock ISBN 9781450335317.
\newblock \doi{10.1145/2882903.2899401}.
\newblock URL \url{https://doi.org/10.1145/2882903.2899401}.

\bibitem[Clyburne-Sherin et~al.(2019)Clyburne-Sherin, Fei, and
  Green]{clyburne2019computational}
April Clyburne-Sherin, Xu~Fei, and Seth Green.
\newblock Computational reproducibility via containers in psychology.
\newblock \emph{Meta-Psychology}, 3, 11 2019.
\newblock \doi{10.15626/MP.2018.892}.

\bibitem[Cohen-Boulakia et~al.(2017)Cohen-Boulakia, Belhajjame, Collin,
  Chopard, Froidevaux, Gaignard, Hinsen, Larmande, Bras, Lemoine, Mareuil,
  Ménager, Pradal, and Blanchet]{Sarah2017}
Sarah Cohen-Boulakia, Khalid Belhajjame, Olivier Collin, Jérôme Chopard,
  Christine Froidevaux, Alban Gaignard, Konrad Hinsen, Pierre Larmande, Yvan~Le
  Bras, Frédéric Lemoine, Fabien Mareuil, Hervé Ménager, Christophe Pradal,
  and Christophe Blanchet.
\newblock Scientific workflows for computational reproducibility in the life
  sciences: Status, challenges and opportunities.
\newblock \emph{Future Generation Computer Systems}, 75:\penalty0 284--298,
  2017.
\newblock ISSN 0167-739X.
\newblock \doi{10.1016/j.future.2017.01.012}.
\newblock URL \url{https://doi.org/10.1016/j.future.2017.01.012}.

\bibitem[Collins et~al.(2023)Collins, Smart, Albright, and Crippin]{e29}
Stephen Collins, Gavin Smart, Ben Albright, and David Crippin.
\newblock Food inspections evaluation.
\newblock url{https://chicago.github.io/food-inspections-evaluation/}, May
  2023.

\bibitem[Cook and Campbell(1979)]{Cook1979}
{Thomas D} Cook and {D T} Campbell.
\newblock \emph{Quasi-Experimentation: Design and Analysis Issues for Field
  Settings}.
\newblock Houghton Mifflin, Boston, 1979.

\bibitem[Costa(2022{\natexlab{a}})]{costa2022platform}
L{\'a}zaro Costa.
\newblock A platform for the reproducibility of computational experiments.
\newblock In \emph{2022 IEEE Symposium on Visual Languages and Human-Centric
  Computing (VL/HCC)}, pages 1--3, Rome, Italy, 2022{\natexlab{a}}. IEEE.
\newblock \doi{10.1109/VL/HCC53370.2022.9833132}.

\bibitem[Costa and da~Silva(2019)]{LCosta2019}
L{\'a}zaro Costa and Jo{\~a}o~Rocha da~Silva.
\newblock Dendro: A fair, open-source data sharing platform.
\newblock In Antoine Doucet, Antoine Isaac, Koraljka Golub, Trond Aalberg, and
  Adam Jatowt, editors, \emph{Digital Libraries for Open Knowledge}, pages
  384--387, Cham, 2019. Springer International Publishing.
\newblock ISBN 978-3-030-30760-8.
\newblock \doi{10.1007/978-3-030-30760-8_39}.
\newblock URL \url{https://doi.org/10.1007/978-3-030-30760-8_39}.

\bibitem[Costa(2022{\natexlab{b}})]{vlhcc2022}
Lázaro Costa.
\newblock A platform for the reproducibility of computational experiments.
\newblock In \emph{2022 IEEE Symposium on Visual Languages and Human-Centric
  Computing (VL/HCC)}, pages 1--3, 2022{\natexlab{b}}.
\newblock \doi{10.1109/VL/HCC53370.2022.9833132}.

\bibitem[Coupette et~al.(2022)Coupette, Hartung, Beckedorf, Böther, and
  Katz]{Coupette2022}
Corinna Coupette, Dirk Hartung, Janis Beckedorf, Maximilian Böther, and
  Daniel~Martin Katz.
\newblock Law smells (code), April 2022.
\newblock URL \url{https://doi.org/10.5281/zenodo.6468193}.

\bibitem[Edwards(2022)]{edwards2022}
Brandon~PM Edwards.
\newblock {napops: Point Count Offsets for Population Sizes of North American
  Landbirds}, February 2022.
\newblock URL \url{https://doi.org/10.5281/zenodo.5967578}.

\bibitem[Erickson(2019)]{Erickson2019}
Adam Erickson.
\newblock adam-erickson/deepland: Deepland alpha release, December 2019.
\newblock URL \url{https://doi.org/10.5281/zenodo.3568468}.

\bibitem[Essawy(2023)]{e30}
Bakinam Essawy.
\newblock Vic\_pre-processing\_rules.
\newblock
  url{https://github.com/uva-hydroinformatics/VIC\_Pre-Processing\_Rules}, May
  2023.

\bibitem[Freire and Silva(2012)]{Freire2012}
Juliana Freire and Claudio~T. Silva.
\newblock Making computations and publications reproducible with vistrails.
\newblock \emph{Computing in Science \& Engineering}, 14\penalty0 (4):\penalty0
  18--25, 2012.
\newblock \doi{10.1109/MCSE.2012.76}.

\bibitem[Freire et~al.(2012)Freire, Bonnet, and Shasha]{Freire2012_b}
Juliana Freire, Philippe Bonnet, and Dennis Shasha.
\newblock Computational reproducibility: State-of-the-art, challenges, and
  database research opportunities.
\newblock In \emph{Proceedings of the 2012 ACM SIGMOD International Conference
  on Management of Data}, SIGMOD '12, page 593–596, New York, NY, USA, 2012.
  Association for Computing Machinery.
\newblock ISBN 9781450312479.
\newblock \doi{10.1145/2213836.2213908}.
\newblock URL \url{https://doi.org/10.1145/2213836.2213908}.

\bibitem[Fries(2021)]{Frie2021}
Jason~Alan Fries.
\newblock som-shahlab/trove: Manuscript pre-release code, February 2021.
\newblock URL \url{https://doi.org/10.5281/zenodo.4497214}.

\bibitem[Hai(2023)]{e31}
Ton Hai.
\newblock Iqe.
\newblock url{https://bitbucket.org/TonHai/iqe/src/master/}, May 2023.

\bibitem[Han(2021)]{Tianyu2021}
Tianyu Han.
\newblock {peterhan91/Medical-Robust-Training: Adversarial Training on Medical
  Data}, June 2021.
\newblock URL \url{https://doi.org/10.5281/zenodo.4926118}.

\bibitem[Hemes(2023)]{Hemes2023}
Kyle Hemes.
\newblock {Data: The magnitude and pace of photosynthetic recovery after
  wildfire in California ecosystems}, April 2023.
\newblock URL \url{https://doi.org/10.5281/zenodo.7651416}.
\newblock {Funding provided by: California Strategic Growth Council's Climate
  Change Research Program* Crossref Funder Registry ID: Award Number:
  CCR20021}.

\bibitem[Hoyt(2018)]{Hoyt2018}
Charles~Tapley Hoyt.
\newblock bio2bel/mesh v0.2.0, November 2018.
\newblock URL \url{https://doi.org/10.5281/zenodo.1488650}.

\bibitem[Huang and Wu(2018)]{Dijiang2018}
Dijiang Huang and Huijun Wu.
\newblock Chapter 2 - virtualization.
\newblock In Dijiang Huang and Huijun Wu, editors, \emph{Mobile Cloud
  Computing}, pages 31--64. Morgan Kaufmann, 2018.
\newblock ISBN 978-0-12-809641-3.
\newblock \doi{https://doi.org/10.1016/B978-0-12-809641-3.00003-X}.
\newblock URL
  \url{https://www.sciencedirect.com/science/article/pii/B978012809641300003X}.

\bibitem[Ivie and Thain(2018)]{Ivie2018}
Peter Ivie and Douglas Thain.
\newblock Reproducibility in scientific computing.
\newblock \emph{ACM Comput. Surv.}, 51\penalty0 (3), jul 2018.
\newblock ISSN 0360-0300.
\newblock \doi{10.1145/3186266}.
\newblock URL \url{https://doi.org/10.1145/3186266}.

\bibitem[Janin et~al.(2014)Janin, Vincent, and Duraffort]{Janin2014}
Yves Janin, C\'{e}dric Vincent, and R\'{e}mi Duraffort.
\newblock Care, the comprehensive archiver for reproducible execution.
\newblock In \emph{Proceedings of the 1st ACM SIGPLAN Workshop on Reproducible
  Research Methodologies and New Publication Models in Computer Engineering},
  TRUST '14, New York, NY, USA, 2014. Association for Computing Machinery.
\newblock ISBN 9781450329514.
\newblock \doi{10.1145/2618137.2618138}.
\newblock URL \url{https://doi.org/10.1145/2618137.2618138}.

\bibitem[Jehn(2021)]{Jehn2021}
Florian~Ulrich Jehn.
\newblock zutn/extreme-climate-change: Publication ready, March 2021.
\newblock URL \url{https://doi.org/10.5281/zenodo.4588383}.

\bibitem[Kailas et~al.(2022)Kailas, Goto, Homilius, MacRae, and
  Deo]{Kailas2022}
Prajwal Kailas, Shinichi Goto, Max Homilius, Calum~A MacRae, and Rahul~C Deo.
\newblock obi-ml-public/ehr\_deidentification, June 2022.
\newblock URL \url{https://doi.org/10.5281/zenodo.6617957}.

\bibitem[Kim and Nah(2018)]{kim2018internet}
Youngseek Kim and Seungahn Nah.
\newblock Internet researchers' data sharing behaviors.
\newblock \emph{Online Information Review}, 42\penalty0 (1):\penalty0 124--142,
  Jan 2018.
\newblock ISSN 1468-4527.
\newblock \doi{10.1108/OIR-10-2016-0313}.
\newblock URL \url{https://doi.org/10.1108/OIR-10-2016-0313}.

\bibitem[Krieger et~al.(2021)Krieger, Nijzink, Thakur, Ramakrishnan, Roskar,
  and Schymanski]{krieger2021repeatable}
Louis Krieger, Remko Nijzink, Gitanjali Thakur, Chandrasekhar Ramakrishnan, Rok
  Roskar, and Stan Schymanski.
\newblock Repeatable and reproducible workflows using the renku open science
  platform.
\newblock In \emph{EGU General Assembly 2021}, pages EGU21--7655, Online,
  19–30 Apr 2021, 2021.
\newblock \doi{10.5194/egusphere-egu21-7655}.
\newblock URL \url{https://doi.org/10.5194/egusphere-egu21-7655}.

\bibitem[Kukucka et~al.(2022)Kukucka, Pina, Ammann, and Bell]{Kukucka2022}
James Kukucka, Luís Pina, Paul Ammann, and Jonathan Bell.
\newblock Confetti: Amplifying concolic guidance for fuzzers.
\newblock In \emph{2022 IEEE/ACM 44th International Conference on Software
  Engineering (ICSE)}, pages 438--450, Pittsburgh, PA, USA, 2022. IEEE.
\newblock \doi{10.1145/3510003.3510628}.

\bibitem[Langer et~al.(2019)Langer, Bilz, and Gaedke]{Langer2019}
Andr{\'{e}} Langer, Ellen Bilz, and Martin Gaedke.
\newblock Analysis of current {RDM} applications for the interdisciplinary
  publication of research data.
\newblock In Lucie{-}Aim{\'{e}}e Kaffee, Kemele~M. Endris, Maria{-}Esther
  Vidal, Marco Comerio, Mersedeh Sadeghi, David Chaves{-}Fraga, and Pieter
  Colpaert, editors, \emph{Proceedings of the Workshop on Approaches for Making
  Data Interoperable (AMAR) Presented at the SEMANTiCS Conference 2019}, volume
  2447 of \emph{{CEUR} Workshop Proceedings}, Karlsruhe, Germany, September
  2019. CEUR-WS.org.
\newblock URL \url{http://ceur-ws.org/Vol-2447/paper1.pdf}.

\bibitem[Lin(2023)]{Lin2023}
Shuqi Lin.
\newblock {Shuqi-Lin/Dissolved-Oxygen-MLPrediction: Dissolved oxygen and
  hypolimnetic hypoxia predictions via machine learning models in various
  lakes}, February 2023.
\newblock URL \url{https://doi.org/10.5281/zenodo.7613549}.

\bibitem[Liu and Salganik(2019)]{David2019}
David~M. Liu and Matthew~J. Salganik.
\newblock Successes and struggles with computational reproducibility: Lessons
  from the fragile families challenge.
\newblock volume~5, page 2378023119849803, 2019.
\newblock \doi{10.1177/2378023119849803}.
\newblock URL \url{https://doi.org/10.1177/2378023119849803}.

\bibitem[Liu et~al.(2015)Liu, Pacitti, Valduriez, and Mattoso]{liu2015survey}
Ji~Liu, Esther Pacitti, Patrick Valduriez, and Marta Mattoso.
\newblock A survey of data-intensive scientific workflow management.
\newblock \emph{Journal of Grid Computing}, 13\penalty0 (4):\penalty0 457--493,
  2015.
\newblock \doi{10.1007/s10723-015-9329-8}.
\newblock URL \url{https://doi.org/10.1007/s10723-015-9329-8}.

\bibitem[Meng et~al.(2015)Meng, Kommineni, Pham, Gardner, Malik, and
  Thain]{Haiyan2015}
Haiyan Meng, Rupa Kommineni, Quan Pham, Robert Gardner, Tanu Malik, and Douglas
  Thain.
\newblock An invariant framework for conducting reproducible computational
  science.
\newblock \emph{Journal of Computational Science}, 9:\penalty0 137--142, 2015.
\newblock ISSN 1877-7503.
\newblock \doi{https://doi.org/10.1016/j.jocs.2015.04.012}.
\newblock URL
  \url{https://www.sciencedirect.com/science/article/pii/S1877750315000502}.
\newblock Computational Science at the Gates of Nature.

\bibitem[Nguyen and Grunske(2022)]{Nguyen2022}
Hoang~Lam Nguyen and Lars Grunske.
\newblock Bedivfuzz: Integrating behavioral diversity into generator-based
  fuzzing.
\newblock In \emph{2022 IEEE/ACM 44th International Conference on Software
  Engineering (ICSE)}, pages 249--261, Pittsburgh, PA, USA, 2022. IEEE.
\newblock \doi{10.1145/3510003.3510182}.

\bibitem[N{\"{u}}st and Pebesma(2021)]{Nust2021}
Daniel N{\"{u}}st and Edzer Pebesma.
\newblock {Practical Reproducibility in Geography and Geosciences}.
\newblock \emph{Annals of the American Association of Geographers},
  111\penalty0 (5):\penalty0 1300--1310, 2021.
\newblock ISSN 24694460.
\newblock \doi{10.1080/24694452.2020.1806028}.

\bibitem[Pala and García-Cuesta(2023)]{Pala2023}
Riccardo Pala and Esteban García-Cuesta.
\newblock {AXOM - Combination of Weak Learners eXplanations to Improve Random
  Forest eXplicability Robustness (Source Code)}, January 2023.
\newblock URL \url{https://doi.org/10.5281/zenodo.7585845}.
\newblock None.

\bibitem[Pasquier et~al.(2018)Pasquier, Lau, Han, Fong, Lerner, Boose, Crosas,
  Ellison, and Seltzer]{Pasquier2018}
Thomas Pasquier, Matthew~K. Lau, Xueyuan Han, Elizabeth Fong, Barbara~S.
  Lerner, Emery~R. Boose, Mercè Crosas, Aaron~M. Ellison, and Margo Seltzer.
\newblock Sharing and preserving computational analyses for posterity with
  encapsulator.
\newblock \emph{Computing in Science \& Engineering}, 20\penalty0 (4):\penalty0
  111--124, 2018.
\newblock \doi{10.1109/MCSE.2018.042781334}.

\bibitem[Pham et~al.(2015)Pham, Malik, and Foster]{Pham2015}
Quan Pham, Tanu Malik, and Ian Foster.
\newblock Auditing and maintaining provenance in software packages.
\newblock In Bertram Lud{\"a}scher and Beth Plale, editors, \emph{Provenance
  and Annotation of Data and Processes}, pages 97--109, Cham, 2015. Springer
  International Publishing.
\newblock ISBN 978-3-319-16462-5.
\newblock \doi{https://doi.org/10.1007/978-3-319-16462-5_8}.

\bibitem[Potdar et~al.(2020)Potdar, {D G}, Kengond, and Mulla]{Amit2020}
Amit~M Potdar, Narayan {D G}, Shivaraj Kengond, and Mohammed~Moin Mulla.
\newblock Performance evaluation of docker container and virtual machine.
\newblock \emph{Procedia Computer Science}, 171:\penalty0 1419--1428, 2020.
\newblock ISSN 1877-0509.
\newblock \doi{10.1016/j.procs.2020.04.152}.
\newblock URL \url{https://doi.org/10.1016/j.procs.2020.04.152}.
\newblock Third International Conference on Computing and Network
  Communications (CoCoNet'19).

\bibitem[Prezja(2023)]{Prezja2013}
Fabi Prezja.
\newblock {Deep Fast Vision: Accelerated Deep Transfer Learning Vision
  Prototyping and Beyond}, April 2023.
\newblock URL \url{https://doi.org/10.5281/zenodo.7867100}.

\bibitem[Qasmi and Ribes(2021)]{Qasmi2021}
Saïd Qasmi and Aurélien Ribes.
\newblock Kriging for climate change package, August 2021.
\newblock URL \url{https://doi.org/10.5281/zenodo.5233947}.

\bibitem[Santucci(2023)]{Santucci2023}
Valentino Santucci.
\newblock valentinosantucci/hydrojump: Evoapps2023, February 2023.
\newblock URL \url{https://doi.org/10.5281/zenodo.7595510}.

\bibitem[Shevchenko(2022)]{Shevchenko2022}
Yury Shevchenko.
\newblock Open lab: A web application for running and sharing online
  experiments.
\newblock \emph{Behavior Research Methods}, 54\penalty0 (6):\penalty0
  3118--3125, Dec 2022.
\newblock ISSN 1554-3528.
\newblock \doi{10.3758/s13428-021-01776-2}.
\newblock URL \url{https://doi.org/10.3758/s13428-021-01776-2}.

\bibitem[Stodden et~al.(2012)Stodden, Hurlin, and Pérignon]{Stodden2012}
Victoria Stodden, Christophe Hurlin, and Christophe Pérignon.
\newblock Runmycode.org: A novel dissemination and collaboration platform for
  executing published computational results.
\newblock In \emph{2012 IEEE 8th International Conference on E-Science}, pages
  1--8, Chicago, IL, USA, 2012. IEEE.
\newblock \doi{10.1109/eScience.2012.6404455}.

\bibitem[Stodden et~al.(2016)Stodden, McNutt, Bailey, Deelman, Gil, Hanson,
  Heroux, Ioannidis, and Taufer]{Victoria2016}
Victoria Stodden, Marcia McNutt, David~H. Bailey, Ewa Deelman, Yolanda Gil,
  Brooks Hanson, Michael~A. Heroux, John~P.A. Ioannidis, and Michela Taufer.
\newblock Enhancing reproducibility for computational methods.
\newblock \emph{Science}, 354\penalty0 (6317):\penalty0 1240--1241, 2016.
\newblock \doi{10.1126/science.aah6168}.
\newblock URL \url{https://www.science.org/doi/abs/10.1126/science.aah6168}.

\bibitem[Sun et~al.(2022)Sun, Zhang, Sun, Li, and Tang]{Sun2022}
Ji~Sun, Jintao Zhang, Zhaoyan Sun, Guoliang Li, and Nan Tang.
\newblock Learned cardinality estimation: A design space exploration and a
  comparative evaluation, jan 2022.
\newblock ISSN 2150-8097.
\newblock URL \url{https://doi.org/10.14778/3485450.3485459}.

\bibitem[Tovar et~al.(2023)Tovar, Thain, Shaffer, and Rajan]{e26}
Benjamin Tovar, Douglas Thain, Tim Shaffer, and Dinesh Rajan.
\newblock Cooperative computing tools (cctools).
\newblock url{https://github.com/cooperative-computing-lab/cctools}, May 2023.

\bibitem[Trueto(2021)]{Trueto2021}
Trueto.
\newblock {trueto/transformers\_sklearn: transformers-sklearn V1.1}, January
  2021.
\newblock URL \url{https://doi.org/10.5281/zenodo.4453803}.

\bibitem[van Hee(2002)]{aslst2002workflow}
Kees van Hee.
\newblock \emph{{Workflow Management: Models, Methods, and Systems}}.
\newblock The MIT Press, 01 2002.
\newblock ISBN 9780262285414.
\newblock \doi{10.7551/mitpress/7301.001.0001}.
\newblock URL \url{https://doi.org/10.7551/mitpress/7301.001.0001}.

\bibitem[Wohlin et~al.(2012)Wohlin, Runeson, Höst, Ohlsson, Regnell, and
  Wesslén]{Wohlin2012}
Claes Wohlin, Per Runeson, Martin Höst, Magnus~C. Ohlsson, Björn Regnell, and
  Anders Wesslén.
\newblock \emph{Experimentation in software engineering}, volume 9783642290442.
\newblock Springer-Verlag Berlin Heidelberg, 2012.
\newblock ISBN 978-364229044-2; 3642290434; 978-364229043-5.
\newblock \doi{10.1007/978-3-642-29044-2}.
\newblock URL \url{https://link.springer.com/book/10.1007/978-3-642-29044-2}.

\bibitem[Xie et~al.(2022)Xie, Lei, Yan, Yu, Xia, and Mao]{Xie2022}
Huan Xie, Yan Lei, Meng Yan, Yue Yu, Xin Xia, and Xiaoguang Mao.
\newblock A universal data augmentation approach for fault localization.
\newblock In \emph{Proceedings of the 44th International Conference on Software
  Engineering}, ICSE '22, page 48–60, New York, NY, USA, 2022. Association
  for Computing Machinery.
\newblock ISBN 9781450392211.
\newblock \doi{10.1145/3510003.3510136}.
\newblock URL \url{https://doi.org/10.1145/3510003.3510136}.

\bibitem[Youngdahl et~al.(2019)Youngdahl, Ton-That, and Malik]{Youngdahl2019}
Andrew Youngdahl, Dai-Hai Ton-That, and Tanu Malik.
\newblock Sciinc: A container runtime for incremental recomputation.
\newblock In \emph{2019 15th International Conference on eScience (eScience)},
  pages 291--300, San Diego, CA, USA, 2019. IEEE.
\newblock \doi{10.1109/eScience.2019.00040}.

\bibitem[Zhang et~al.(2022)Zhang, Chen, Peng, and Zhao]{Zhang2022}
Chen Zhang, Bihuan Chen, Xin Peng, and Wenyun Zhao.
\newblock Buildsheriff: Change-aware test failure triage for continuous
  integration builds.
\newblock In \emph{2022 IEEE/ACM 44th International Conference on Software
  Engineering (ICSE)}, pages 312--324, Pittsburgh, PA, USA, 2022. IEEE.
\newblock \doi{10.1145/3510003.3510132}.

\bibitem[Zheng(2021)]{Zuduo2021}
Zuduo Zheng.
\newblock Reasons, challenges, and some tools for doing reproducible
  transportation research.
\newblock \emph{Communications in Transportation Research}, 1:\penalty0 100004,
  2021.
\newblock ISSN 2772-4247.
\newblock \doi{https://doi.org/10.1016/j.commtr.2021.100004}.
\newblock URL
  \url{https://www.sciencedirect.com/science/article/pii/S2772424721000044}.

\bibitem[Zhou et~al.(2022)Zhou, Wang, and Chen]{Zhou2022}
Alexander Zhou, Yue Wang, and Lei Chen.
\newblock Butterfly counting on uncertain bipartite graphs.
\newblock \emph{Proc. VLDB Endow.}, 15\penalty0 (2):\penalty0 211–223, feb
  2022.
\newblock ISSN 2150-8097.
\newblock \doi{10.14778/3489496.3489502}.
\newblock URL \url{https://doi.org/10.14778/3489496.3489502}.

\end{thebibliography}

%----------------------------------------------------------------------------------------

\end{document}